\documentclass[useAMS,usenatbib,usegraphicx]{mn2e}
\usepackage{aas_macros,amssymb,amsmath}
\usepackage{url}

\newcommand{\ud}{{\rm d}}
\newcommand{\gsim}{\;\rlap{\lower 2.5pt
 \hbox{$\sim$}}\raise 1.5pt\hbox{$>$}\;}
\newcommand{\lsim}{\;\rlap{\lower 2.5pt
   \hbox{$\sim$}}\raise 1.5pt\hbox{$<$}\;}
\newcommand{\be}{\begin{equation}}
\newcommand{\beq}{\begin{equation}}
\newcommand{\ba}{\begin{eqnarray}}
\newcommand{\ee}{\end{equation}}
\newcommand{\eeq}{\end{equation}}
\newcommand{\ea}{\end{eqnarray}}

\newcommand{\bea}{\begin{eqnarray}}
\newcommand{\eea}{\end{eqnarray}}
\newcommand{\bean}{\begin{eqnarray*}}
\newcommand{\eean}{\end{eqnarray*}}

\newcommand{\bx}{{\bf x}}

\newcommand{\bk}{{\bf k}}

\newcommand{\tcmb}{T_\gamma}
\newcommand{\Lyman}{${\rm Ly\alpha}$ }

\begin{document}

\title[Fast Large Volume Simulations of the 21 cm signal]{Fast Large Volume Simulations of the 21 cm Signal from the Reionization and pre-Reionization Epochs}

\author[M. G. Santos et al.]
{M. G. Santos$^{1}$\thanks{Email: mgrsantos@ist.utl.pt (MGS); luis.ferramacho@ist.utl.pt (LF)}, L. Ferramacho$^{1}$, M. B. Silva$^{1}$,
A. Amblard$^{2}$ and A. Cooray$^{2}$\\
$^{1}$ CENTRA, Departamento de F\'isica, Instituto Superior T\'ecnico, 1049-001 Lisboa, Portugal\\
$^{2}$ Center for Cosmology, Department of Physics and Astronomy, University of California, Irvine, CA 92697}

\maketitle

\begin{abstract}

While limited to low spatial resolution, the next generation low-frequency radio interferometers that target 21 cm observations during the era of reionization
and prior will have instantaneous fields-of-view that are many tens of square degrees on the sky. Predictions related to various statistical measurements of the 
21 cm brightness temperature must then be pursued 
with numerical simulations of reionization with correspondingly large volume box sizes, of order 1000 Mpc on one side. 
We pursue a semi-numerical scheme to simulate the 21 cm signal during and prior to Reionization by extending a hybrid approach where simulations are performed
by first laying down the linear dark matter density field, accounting for the non-linear evolution of the density field based on
second-order linear perturbation theory as specified by the Zel'dovich approximation,  
and then specifying the location and mass of collapsed dark matter halos using the excursion-set formalism. The location of ionizing sources
and the time evolving distribution of ionization field is also specified using an excursion-set algorithm. We account for the brightness temperature evolution through the coupling between spin and gas temperature due to collisions, 
radiative coupling in the presence of Lyman-alpha photons  and heating of the intergalactic medium, such as due to a background of
X-ray photons. The hybrid simulation method we present is capable of producing the required large volume simulations with adequate resolution in a reasonable time
so a large number of realizations can be obtained
with variations in assumptions related to astrophysics and background cosmology that govern the 21 cm signal.

\end{abstract}


\section{Introduction}

Observations of the 21 cm line of neutral hydrogen are currently considered to be one of the most promising probes of the epoch of reionization (EoR) and possibly even the preceding period, during the so called dark ages \citep{madau97, loeb04, gnedin04, furlanetto04a, zaldarriaga04,sethi05, bharadwaj05, morales03, loeb04}. 
Given the line emission, leading to a frequency selection for observations, the 21 cm data provides a tomographic view of the reionization and pre-reionization process \citep{santos05, furlanetto04c} and can also be an exquisite cosmological probe providing complimentary information when compared to cosmic
microwave background anisotropies \citep{mao08,santos06,mcquinn06,bowman07}.
New large area radio interferometers are now being developed, with the promise of measuring this signal, namely, LOFAR\footnote{http://lofar.org} in the Netherlands, MWA\footnote{http://web.haystack.mit.edu/arrays/MWA} in Australia and the low-frequency extension of the 
planned Square Kilometre Array (SKA\footnote{http://www.skatelescope.org}).

Motivated by the observational possibilities offered by the current and upcoming low frequency radio interferometers a great deal of effort has been underway 
in order to fully understand and generate the expected 21 cm signal that will be seen by these experiments (see \citealt{furlanetto06c} for a review). 
Analytical models can be very useful to quickly generate the signal with high dynamic range, in particular the power spectrum of 21 cm brightness temperature
fluctuations \citep{furlanetto04b,furlanetto06a,sethi05,barkana07} and for testing the ability of a given experiment to constraint cosmological and
astrophysical parameters \citep{santos06,mcquinn06,mao08}. The analytical models have also been useful to understand the possible contributions to the 21 cm signal at high redshifts \citep{barkana05,pritchard06}, although it is at low redshifts ($z\lesssim 10$) that they seem to provide a better description of the 21 cm 2-point correlation function \citep{santos08}. However, these models still have problems dealing with the spatial distribution of the reionization process, such as bubble overlap
and ignores complicated astrophysics during reionization.

Numerical simulations, on the other hand, can potentially provide an improved description of the 21 cm brightness temperature signal from first principles. 
They typically involve a combination of a N-body algorithm for dark matter, coupled to a prescription for baryons and star formation (often complemented by a hydrodynamical simulation), plus a full radiative transfer code to propagate the ionizing radiation \citep{gnedin00a, razoumov02,sokasian03,ciardi03,kohler05,iliev06,zahn06,mcquinn07,shin07,trac06,baek09}. These simulations have the advantage of properly dealing with the spatial distribution of the fields (such as bubble overlapping or the non-sphericity of the bubbles) and can be used to extract more information than just the 21 cm anisotropy power spectrum.
The disadvantage is that they are slow to run, being constrained in dynamical range to sizes typically smaller than 143 Mpc,
and producing a large set of simulations for cosmological and astrophysical parameter estimates is unrealistic.

Instead the 21 cm community makes use of hybrid approaches involving simulations and analytical techniques.
\citet{thomas09} generate faster simulations by post-processing a dark matter simulation with a 1-D radiative transfer code to create bubbles around the sources of radiation embedded in the dark matter halos. \cite{mesinger07} have taken a hybrid approach between the analytical models \citep{furlanetto04b,furlanetto04c} and the spatial description provided by the full numerical simulations. This is based on 3-D Monte Carlo realizations of the dark matter density field combined with an ``analytical prescription'' in order to generate the halo and ionization fields. \citet{zahn06} also considered a similar analytical prescription but based directly on the linear density field (see also \citealt{alvarez09}). Although much faster than previous numerical approaches, the hybrid approach of \citet{mesinger07} is still constrained to ``small'' sizes 
depending on the amount of shared memory available in the computer and to low redshifts since it neglects the spin temperature evolution for the 21 cm calculation. 

It is useful to note that the proposed low-frequency interferometers that plan to observe the 21 cm signal from reionization have low spatial resolution capabilities but very large fields of view - 5x5 deg$^2$ or more, requiring boxes of at least 1000 Mpc in size if we want to properly simulate a single field of view and pass such a simulation through the observation pipeline in order to understand how instrumental noise and systematic effects impact the observations. At the same time we need to be able to resolve halos of the order of 10$^8 M_\odot$ corresponding to a cooling temperature of $10^4$ K that is deemed necessary to host the ionizing sources responsible for reionization. This implies a resolution of 0.14 Mpc requiring the use of large boxes with $(7000)^3$ cells, making even a hybrid method not useful. Moreover, at high redshifts we need to take into account detailed physics that determine the 21 cm brightness temperature, such as the heating of the inter galactic medium (IGM) or the Lyman-$\alpha$ coupling of the spin temperature to the gas temperature, which makes it a challenging exercise to implement a complete radiative transfer/gas dynamics algorithm in a large volume simulation box.

In this paper, we propose a semi-numerical technique, partially following the hybrid approach prescribed in \citet{mesinger07}, capable of quickly generating an end-to-end simulation of the 21 cm signal even at high redshifts when the effect of the spin temperature is non-negligible. Moreover, this method can be used to simulate very large volumes (e.g. 1000 Mpc), crucial to simulate the field-of-view of next generation of radio telescopes, without sacrificing the speed or requiring unpracticable amounts of computer memory. The code to generate this type of simulation will be provided publicly online\footnote{\url{http://www.simfast21.org}} and it will be subject to continuous improvement through calibration against full radiative transfer/hydrodynamic simulations. The large volume simulations created with this method will be part of the SKA Design Studies Simulated Skies ($S^3$) initiative and we hope this approach can prove useful to generate sky models for the future 21 cm experiments, which is crucial to test for calibration issues and foreground removal as well as to study the features of the 21 cm signal that can be probed by a variety of experiments and to check the dependence of various
statistics extracted from the observations on the underlying astrophysical and cosmological parameters.

In the next Section we introduce the algorithm to generate the simulation of the 21 cm signal up to high redshifts, while in Section~3 we explain how to extend the simulations to very large volumes. For results presented here, we assumed a flat $\Lambda$CDM universe with the following cosmological parameters: $\Omega_m=0.28$, $\Omega_b=0.046$, $h=0.70$, $n_s=0.96$ and $\sigma_8=0.817$ based on the results from WMAP5, BAO and SN (see \citealt{hinshaw09} and references therein). Throughout the paper we quote all quantities in comoving units unless otherwise noted.

\section{Semi-numerical simulation up to {\lowercase{z}}=25}

In this Section, we outline our semi-numerical approach, first focusing on establishing the linear density field and from it the dark matter halo distribution followed by the ionization field, with first order perturbations to account for non-linear effects.

\subsection[]{From linear density fluctuations to the halo distribution}

The underlying basis for our simulation is a Monte-Carlo realization of the dark matter linear density field assuming a Gaussian probability distribution function for the linear overdensity as the initial state, which is then later evolved and used to find the collapsed structures and the corresponding ionized bubbles. This implementation partially follows the algorithm prescribed in \citet{mesinger07}, which is based on the ``peak-patch" approach introduced by \citet{bond96}. The principle behind this procedure is that at the high redshifts such as those where reionization occurs, the spatial distribution of dark matter halos can be obtained with linear theory, and non-linear gravitational effects can be accounted for by using only first-order perturbations on the linear field - i.e., with the Zel'dovich approximation \citep{zeldovich70,efstathiou85}. 

As part of the simulations presented here, we first start by creating a realization of the linear matter density field, using a random Gaussian distribution to generate the Fourier space density fluctuations in a box with comoving size L and N cells:
\begin{equation}
\centering
\delta(\mathbf{k})=\sqrt{\frac{P_{\delta\delta}(k)L^3}{2}}(a_\mathbf{k}+ i b_\mathbf{k})\;,
\label{delta_k}
\end{equation}
where $a_\bk$, $b_\bk$ are Gaussian random variables with unit variance and zero mean and $P_{\delta\delta}(k)$ is the dark matter power spectrum.
We define the power spectrum $P_{aa}(k)$ for a generic quantity $a$, as $\langle a(\bk)a(\bk')^* \rangle=(2\pi)^3 \delta_D^3(\bk-\bk') P_{aa}(k)$ for continuous fields, where $\delta_D$ is the Dirac delta function. 
The power spectrum can be determined from linear evolution of the primordial density fluctuations. Although numerical codes such as CAMB and CMBFAST could be used for a more accurate evaluation of the power spectrum, the simulations presented in this paper use the approximation provided by the \citet{eisenstein98} fitting formulae for the dark matter transfer function in standard $\Lambda$CDM models (note that the our simulation code also includes a routine to read in a transfer function from a Boltzmann equation solver such as CAMB)

The density field in real space then follows from the inverse Fourier transform:
\begin{equation}
\centering
\delta(\mathbf{x})=\frac{1}{L^3}\sum \delta(\mathbf{k})\exp{i\mathbf{k}\cdot\mathbf{x}}\;,
\label{delta_x}
\end{equation}
where $\bx$ and $\bk$ are discretized taking into account the simulation box.
This realization of the dark matter density field can be used to identify collapsed halos using the excursion-set formalism, in which a given region is considered to undergo gravitational collapse if its mean overdensity reaches a certain critical value of order of $\delta_c(z)\sim 1.68/D(z)$, assuming spherical symmetric collapse ($D(z)$ is the linear growth factor normalized so that $D(0)=1$). High resolution N-body simulations  \citep{sheth99,sheth01} showed that by considering ellipticity in non-linear gravitational collapse, the critical overdensity would become a function of the filter scale as:
\begin{equation}
\centering
\delta_c(M,z)=\sqrt{a}\delta_c(z)\left[1+b\left(\frac{\sigma^2(M)}{a\delta^2_c(z)}\right)^c\right]\;,
\label{delta_x2}
\end{equation}
where $\delta_c(z)$ is the quantity introduced above and $a$, $b$ and $c$ are fitting parameters. We have adopted this definition in order to identify dark matter halos. This algorithm simply applies a series of top-hat filters of decreasing size to obtain the mean density around any given point of the density field. 
Note that in order to make the process faster, we applied the filtering procedure as a multiplication in Fourier space (with the corresponding Fourier transform of the window function) and then transformed back to real space.
A given cell is then considered to be the center of a halo of mass $M$ when the condition $\delta(M,x)>\delta_c(M,z)$ is verified. We also consider a full exclusion criterion, which means that halos overlapping with a previously marked halo are discarded. This algorithm allows then to obtain a catalog of halo positions and masses which can be used to identify the star formation regions responsible for ionizing the surrounding IGM. An important factor in this analysis is the resolution used in our linear grid, which sets the minimum size of halos that can be identified. If one wants to go as far as to resolve halos of mass $~10^8M_{\odot}$, we need to have a resolution of order of $0.14$ Mpc. 

The above spatial resolution (which sets the grid size of our simulation box), together with the available RAM memory of the machine used to create these first simulations, 
justifies the maximum size of the simulation to be of L=300 Mpc for N=$1800^3$ cells (resulting in a minimum halo mass of $1.65\times 10^8 M_{\odot}$). For the results presented here we used a machine with 32 CPUs and 64 GB of shared RAM memory, although only 20 CPUs were used at any given time. This algorithm requires the use of large Fast Fourier Transforms (we used the fftw-3.1.1 package\footnote{http://www.fftw.org}) to filter the density field in Fourier space and thus requires a significant amount of allocatable shared memory in order to create simulation boxes with enough resolution. In our machine, we managed to obtain the full halo catalog in a reasonable running time of less than 1 hour per redshift.
   
\begin{figure}
\includegraphics[width=0.5\textwidth]{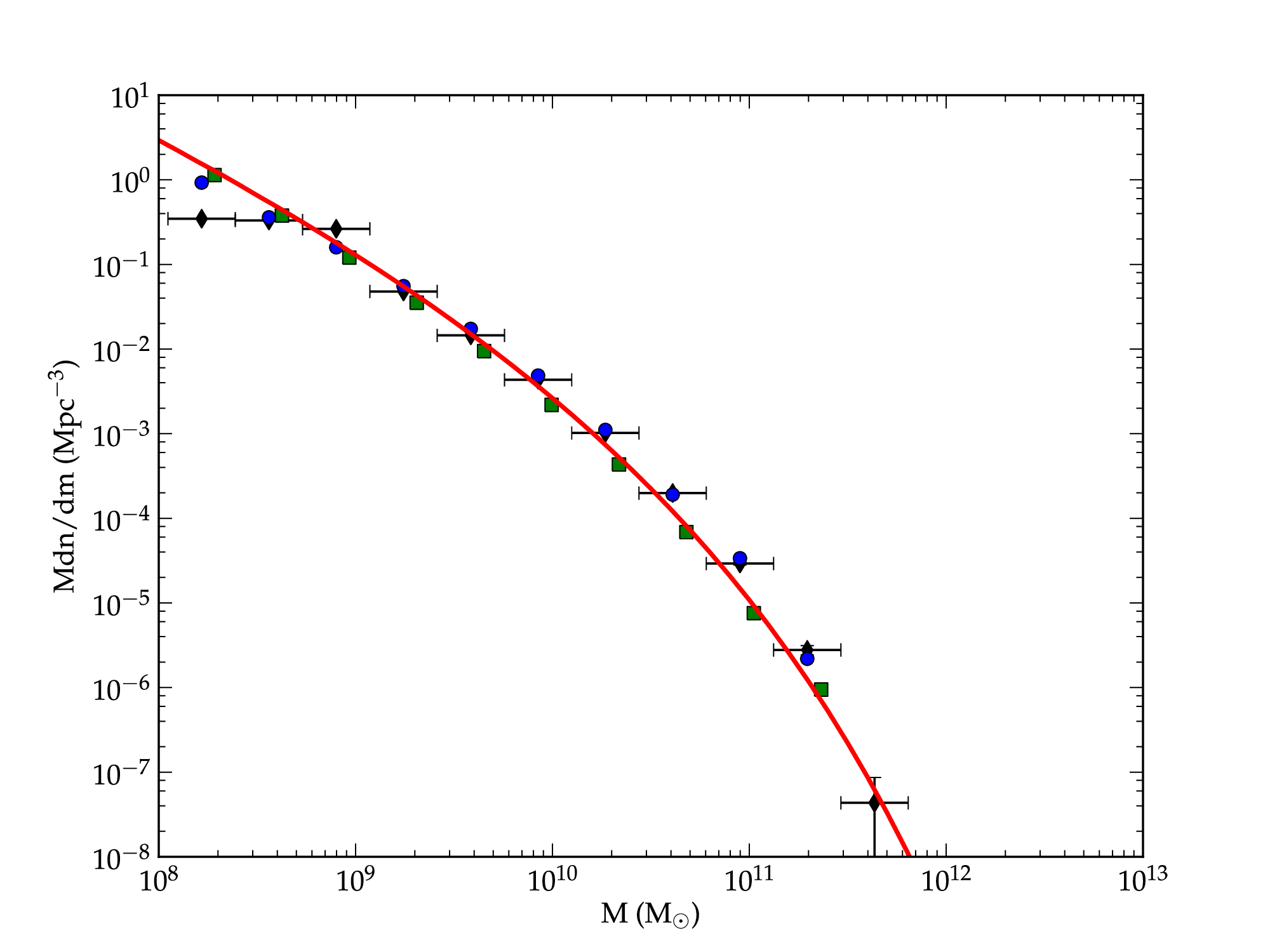}
\caption[]{Mass functions at z=10 taken from our halo filtering prescription with L=143 Mpc and N=1536$^3$  (blue circles) and L=300 Mpc and N=1800$^3$ (black diamonds). The results for the N-body simulation of \citet{trac08} with L=143 Mpc are also shown as green squares. The line show the mass function from \citet{sheth99}, using the fitting parameters by \citet{jenkins01}. Error bars are included only for the L=300Mpc data for viewing purposes.}
\label{fig:mass_func}
\end{figure}

In figure~\ref{fig:mass_func} we show the mass functions obtained at z=10 for two different configurations (L=300 Mpc;N =1800$^3$ and L=143 Mpc;N=1536$^3$) and compare them with the results from the full N-body simulation of \citet{trac08} which covers a region with L=143 Mpc. The straight line represents the expected mass function as derived by \citet{sheth99}, using the fitting parameters by \citet{jenkins01}. One can see that both simulations agree quite well with the theoretical curve although the number of low mass halos in the simulation with the lower resolution (L=300 Mpc) is slightly underestimated. However, we have checked that this effect has a minor impact on the ionized hydrogen distribution and can be seen as a good compromise in order to cover the largest possible volume with this algorithm.    
%%%             
%%%%

With the halo distribution derived from the linear density field, the positions of both the halos and the dark matter were corrected to include non-linear dynamics. This was done by the means of the Zel'dovich approximation, i.e., by using the linear velocity field, obtained from our simulations, to adjust the positions of both the halos and dark matter distribution, starting at an arbitrary high redshift (see \citealt{mesinger07} for details on this implementation). 

\begin{figure}
\includegraphics[width=0.5\textwidth]{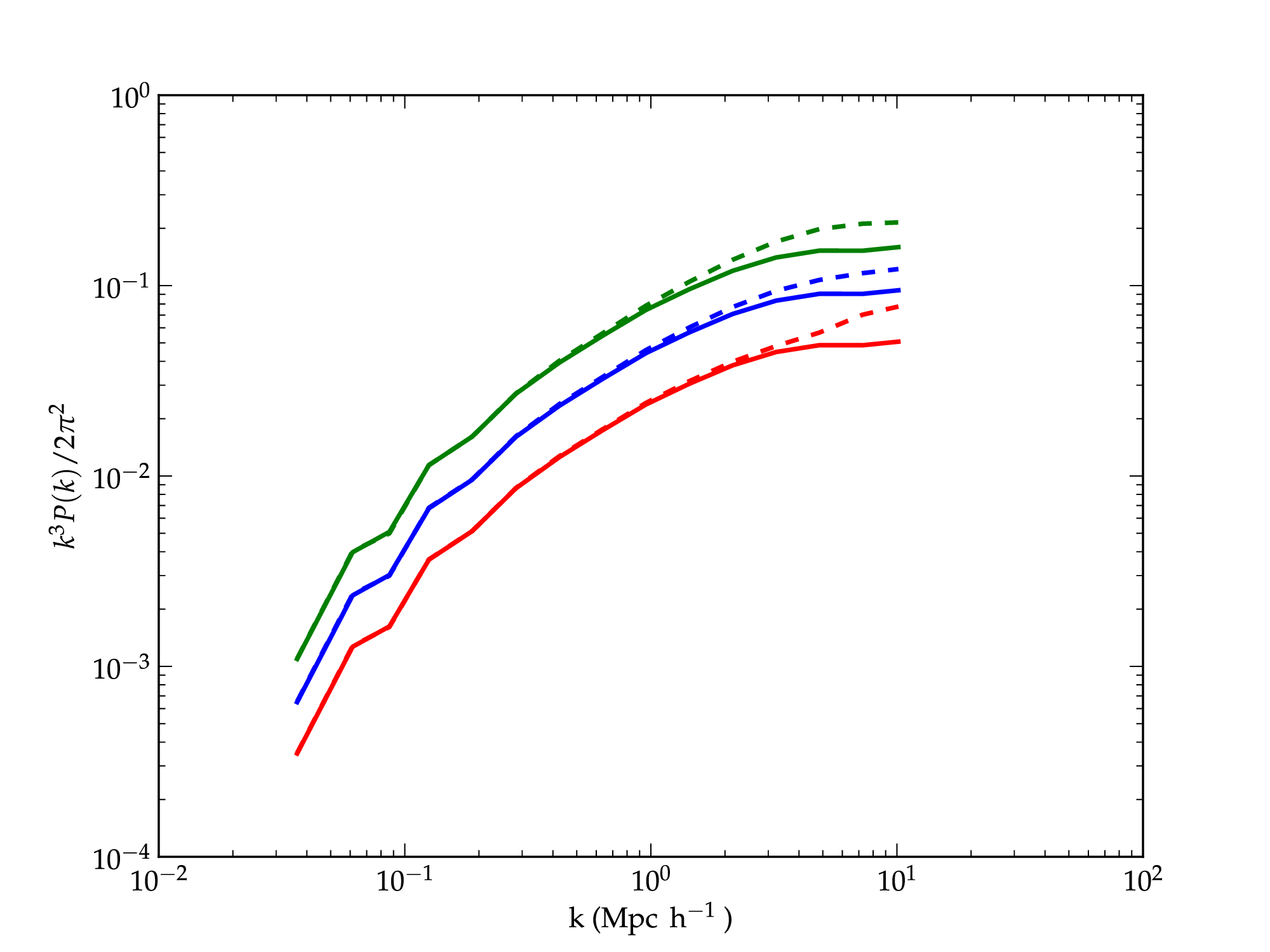}
\caption[]{Normalized power spectrum of matter density fluctuations at redshifts 7.5 (\emph{upper curves}), 10 (\emph{middle curves}) and 14 (\emph{lower curves}). Straight and dashed lines correspond to the power spectrum in our simulation, before and after applying the Zel'dovich corrections.}
\label{fig:pk_delta_nl}
\end{figure}

In figure \ref{fig:pk_delta_nl} we show the power spectrum at three different redshifts of the dark matter fluctuations with and without the non-linear corrections, in the dimensionless form $\Delta(k,z)=k^3 P(k)/2\pi^2)$. In terms of the density power spectrum, we can see that the Zel'dovich approximation increases the correlations at small scales similar to what is seen in full N-body simulations. 
%%%%%%%%%%%%%%%%%%%%%%%

\subsection[]{From the halo distribution to ionization field}

With the corrected halo and density fields, the ionization regions can be determined using a similar excursion-set algorithm. The principle behind this procedure is that the galaxies formed inside dark matter halos will produce a given amount of photons (dependent of the halo mass) that will ionize the surrounding hydrogen generating ionized bubbles. The efficiency of this process can be quantified by one parameter $\zeta$, so that a region will be ionized when:
\begin{equation}
\centering
f_{coll}\geq \zeta^{-1}\ ,
\label{barrier}
\end{equation}
where $f_{coll}$ is the fraction of mass that has collapsed in halos in that region \citep{furlanetto04b}. We assumed a constant efficiency parameter although this can be easily extended to include redshift and mass dependence without degradation of the speed of the simulation. We note that this is an approximate model and contains several limitations when compared with full radiative transfer methods. In particular it does not include radiative feedback effects and spatially dependent recombinations. Neverheless, comparison with such more detailed (albeit slower) simulations show that our bubble filtering algorithm is a very good approximation providing maps with enough accuracy to be compared against future 21cm experiments.

We use this principle to construct our ionization field ($x_i$), averaging both the halo mass and density fields with the spherical top-hat filters of decreasing size. Whenever the condition in equation~(\ref{barrier}) was verified, all cells contained inside the sphere where flagged as completely ionized, allowing at the same time for overlapping regions. To increase the speed of the simulation, we chose to smooth the density and halo fields into a 600$^3$ grid before starting the filtering algorithm. The impact of reducing the resolution is to increase the minimum bubble size, causing some small scale ionization to be lost. To account for this effect we considered a fractional description of the ionization state within each cell. After filtering our box up to the cell size, we searched for non ionized cells containing halos, to which we attributed an ionization value of $x_i=V_i/V_{cell}$, the ratio between the ionized volume which follows directly from our barrier condition $V_i=M_{coll}\zeta/(1+\delta_i)\bar{\rho}_m$ and the cell volume.

Another important aspect of the degrading resolution for ionization calculation can be the overall loss of information on small scales when using smoothed boxes to apply the Zel'dovich approximation. The use of small size boxes is desirable to reduce computational costs, but reducing the box size of the velocity field used to displace halos and dark matter particles can have a significant impact in the final mean ionization fraction ($\bar{x}_i$). We checked that by performing all calculations with smoothed boxes of 300$^3$ or 600$^3$ cells the value of $\bar{x}_i$ can change by up to 5\% for a fixed value of $\zeta$. To keep the maximum of information at all scales we decided to perform all corrections with 1800$^3$ boxes and smooth the density and halo distributions only after applying the Zel'dovich corrections. There is no significant increase in the computational cost when implementing the Zel'dovich approximation on the original box size.

\begin{figure*}
\vspace{+0\baselineskip}
\centerline{
\includegraphics[width=0.55\textwidth]{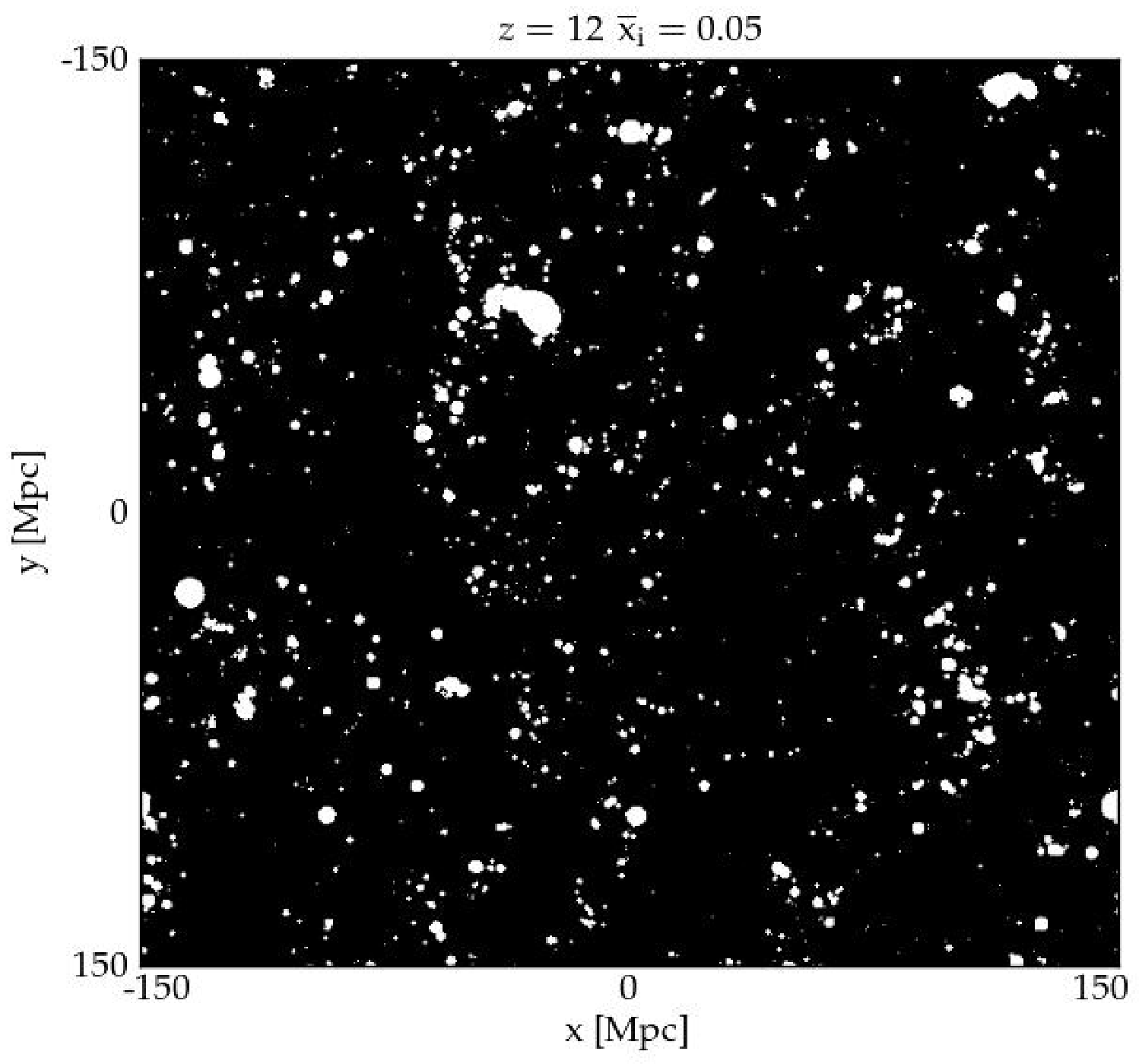}
%\hspace{-2.5cm}
\includegraphics[width=0.55\textwidth]{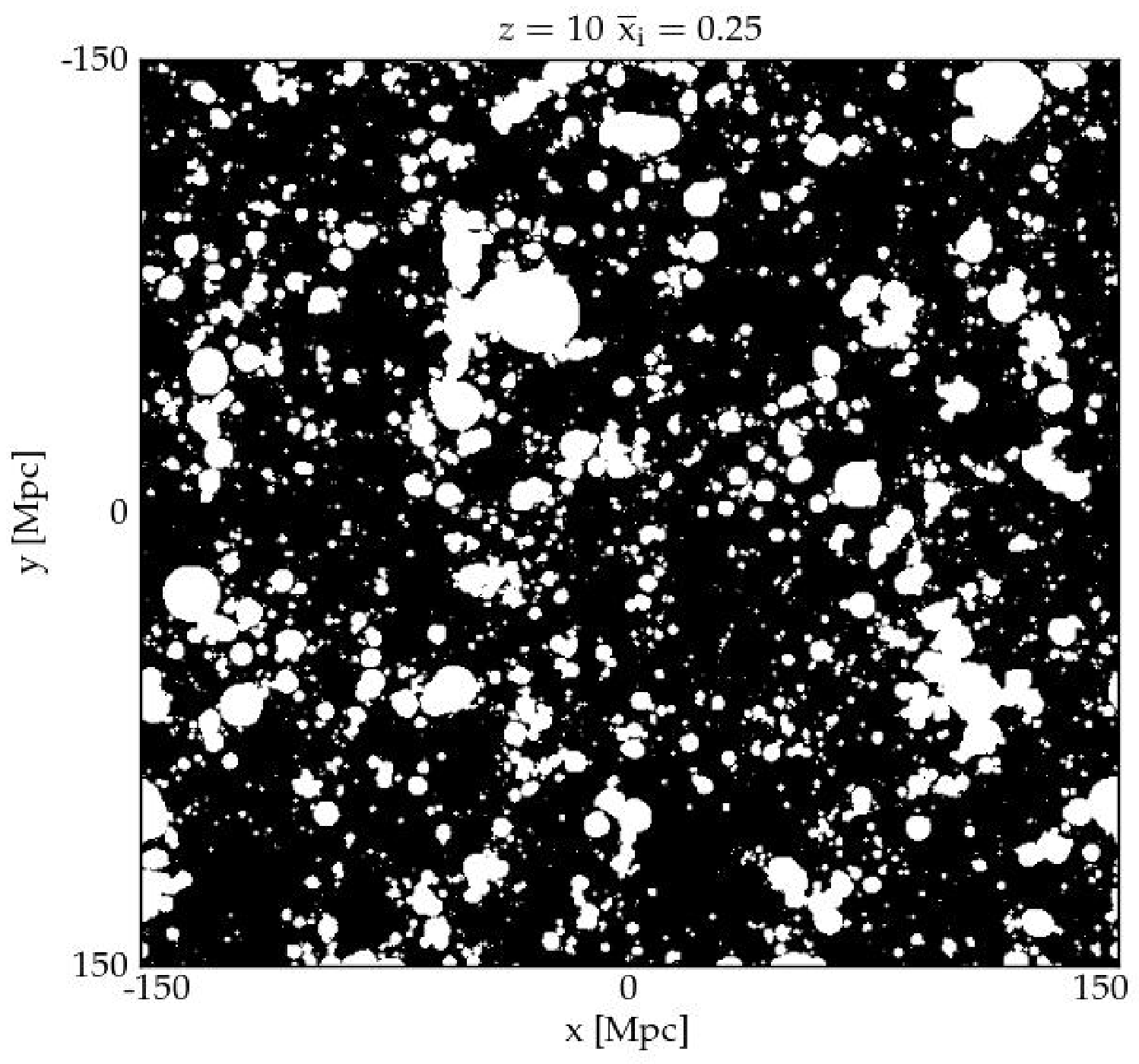}
}
\centerline{
\includegraphics[width=0.55\textwidth]{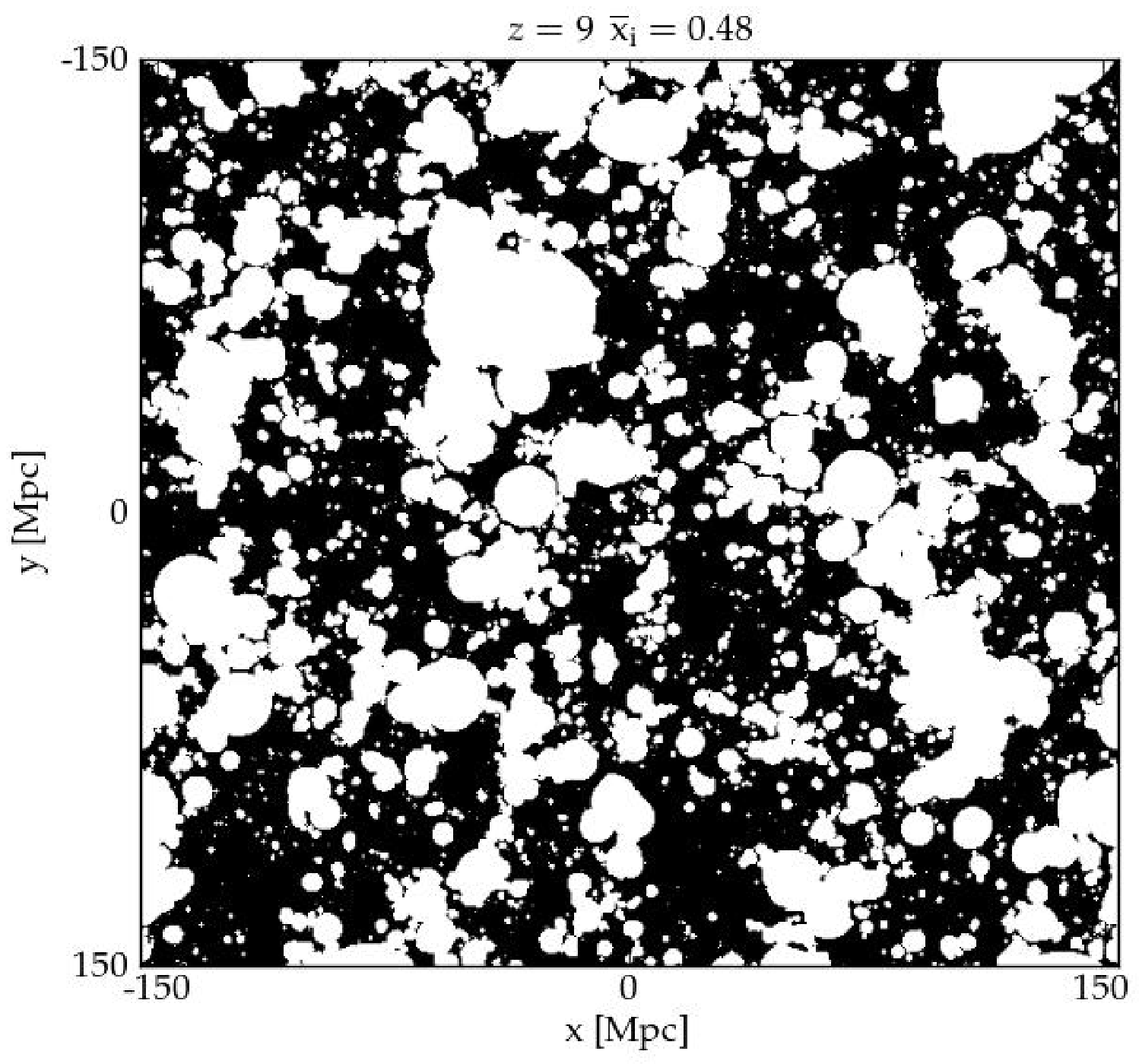}
%\hspace{-2.5cm}
\includegraphics[width=0.55\textwidth]{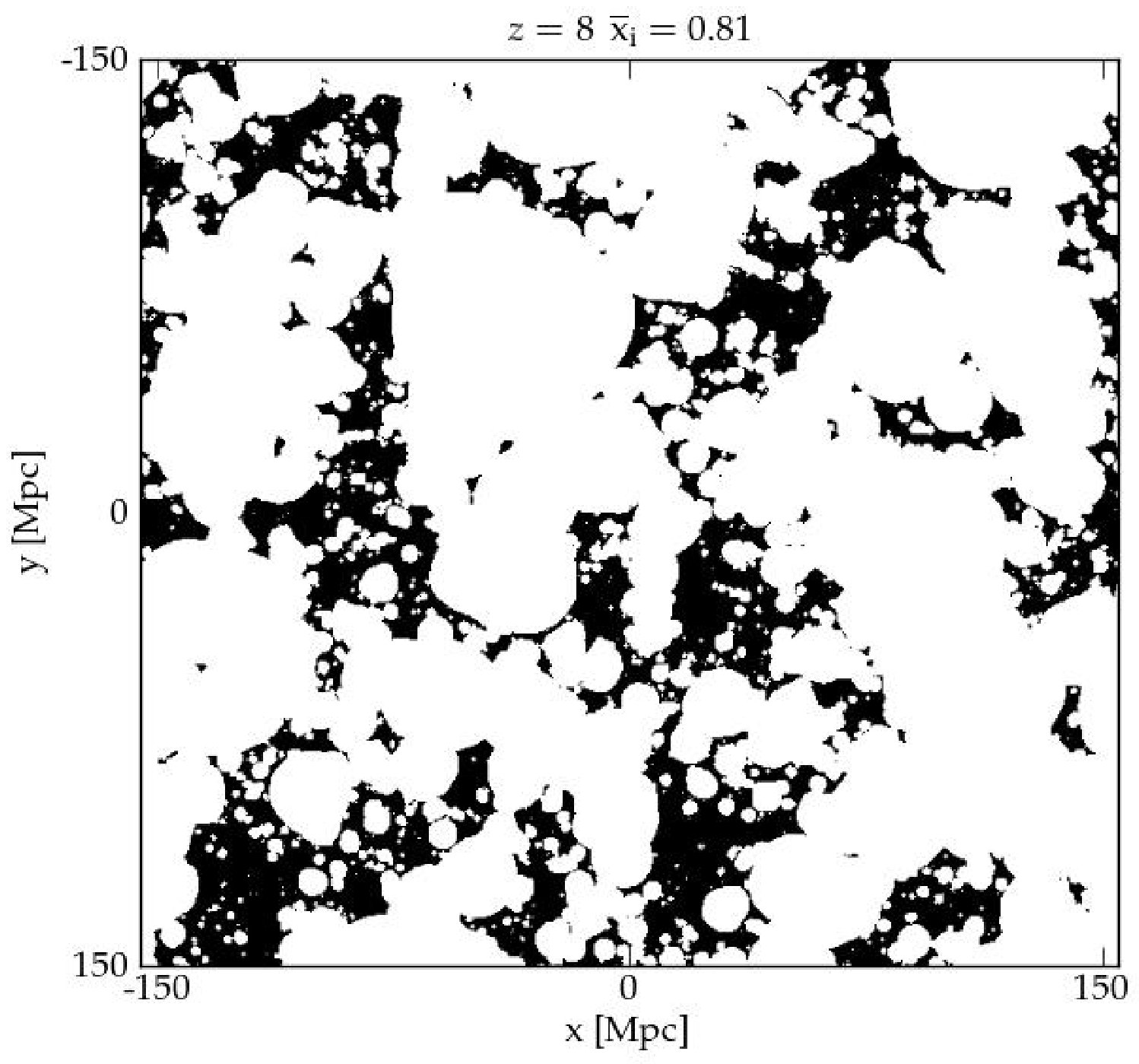}
}
\vspace{-1\baselineskip} \caption[]{
Slices of the ionization field at four different redshifts: clockwise from top-left z = 12, 10, 9, and 8.
\label{fig:bubble_maps}
}
\vspace{-1\baselineskip}
\end{figure*}

\begin{figure}
\includegraphics[width=0.5\textwidth]{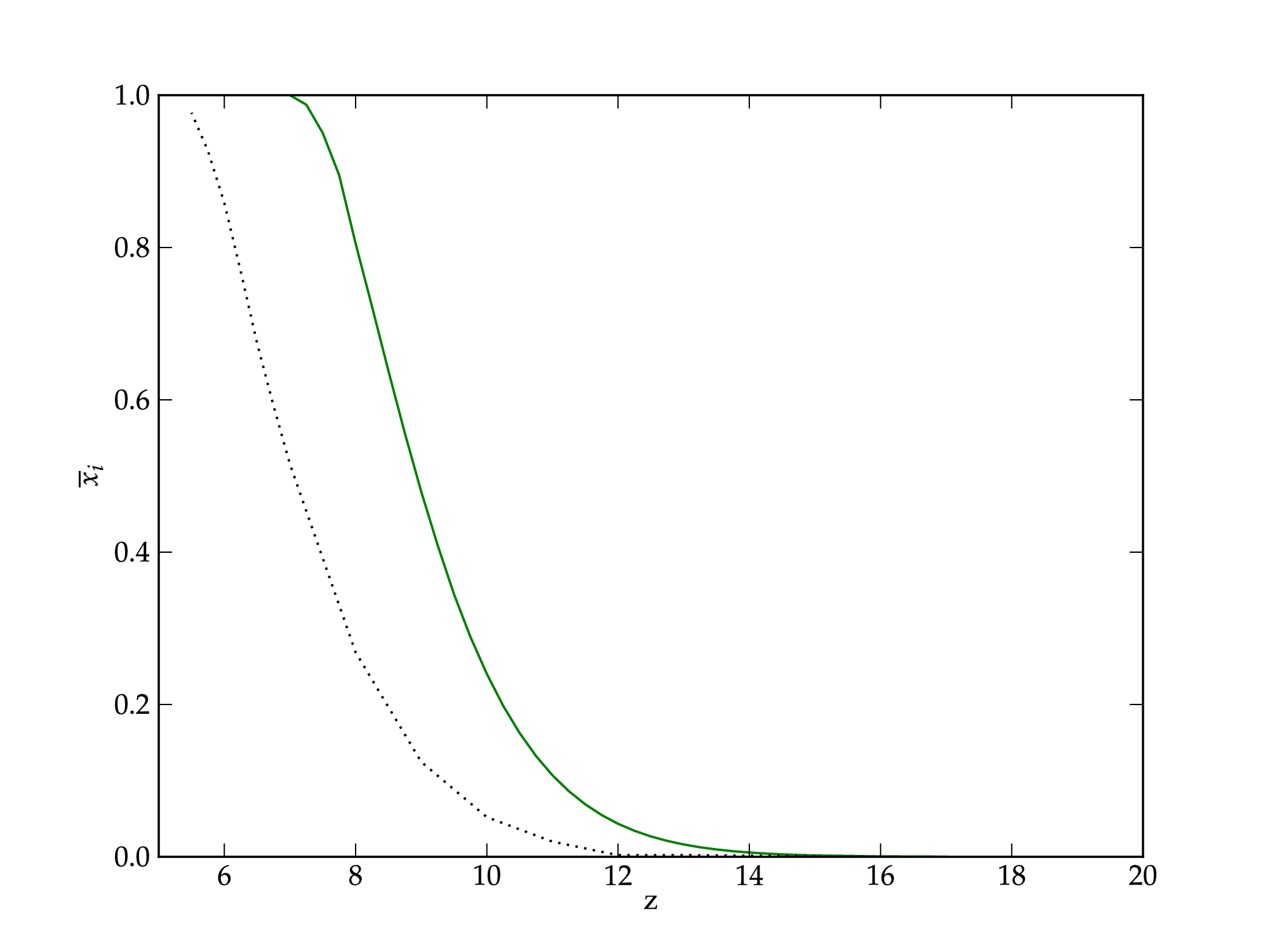}
\caption[]{Evolution of the ionization field $x_i$ using $\zeta=6.7$ (dashed line) and $\zeta=15.1$ (solid line).}
\label{fig_ev}
\end{figure}

As a result of this multi-step algorithm, we obtained the distribution of ionization bubbles in our simulation box which can be used to derive several relevant quantities. In figure~\ref{fig:bubble_maps} are shown 4 slices of our ionization field at decreasing redshifts. These figures exhibit the characteristic bubble structure of ionized hydrogen increasing in size at lower redshifts until converging into a completely ionized IGM. The four panels of figure \ref{fig:bubble_maps} correspond to $\bar{x}_i=$0.08, 0.24, 0.46, 0.82, with the bulk of the reionization happening between z=14 and z=7.5. Obviously, these values are intrinsically dependent on the assumed efficiency (here we assumed a value of 15.1), which could be constrained through observational data. This is shown in figure \ref{fig_ev} where we compare the evolution of the mean ionization fraction with redshift using $\zeta=6.7$ and 15.1. We see that by changing the efficiency parameter we can change the redshift where reionization starts and fit to a desired optical depth (we can quickly generate the ionization field once we change $\zeta$).
%%%%%%%%%%%%%%%%%%%%
%%%%%%%%%%%%%%%%%%%
%%%%%%%%%%%%%%%%%%55

\begin{figure}
\includegraphics[width=0.5\textwidth]{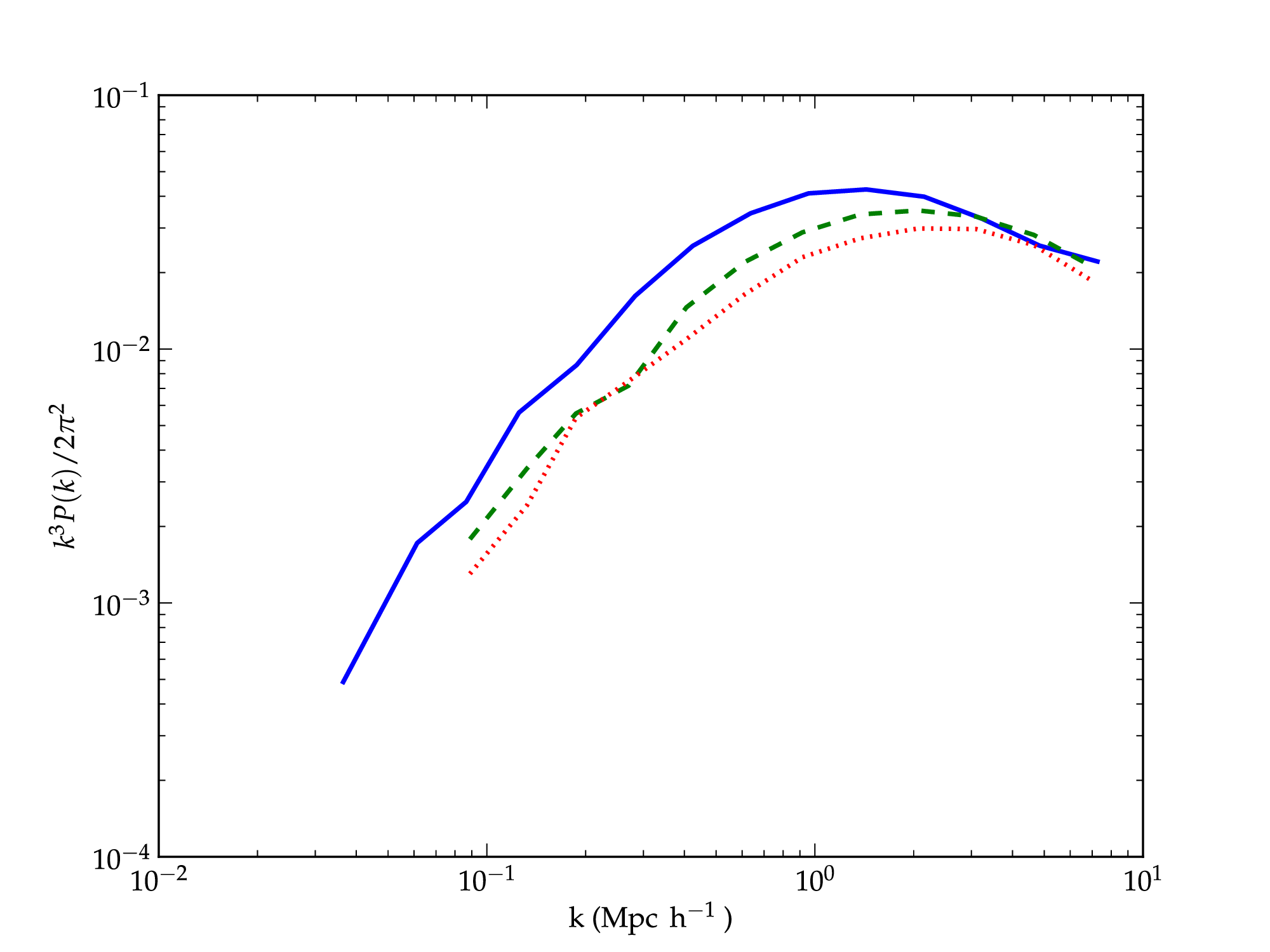}
\caption[]{
Power spectrum of the ionization field for our simulation with L=143 Mpc,N=768$^3$ (red dotted curves), L=300 Mpc,N=600$^3$ (solid blue curves) and the simulation of \citet{trac08} (dashed green curves) for $z=9$ and $\bar{x}_i=0.15$}
\label{pk_bubbles}
\end{figure}

In order to characterize the statistical distribution of the ionization field we computed its power spectrum ($P_{x_ix_i}$). 
In figure~\ref{pk_bubbles} we show the results using two configurations of L=300 Mpc, N=600$^3$ and L=143 Mpc, N=768$^3$ and compare it to the results from \citet{trac08} which also has L=143 Mpc, N=768$^3$. The efficiency parameter was adjusted to obtain the same ionization fraction $\bar{x}_{i}$($\zeta=7$ for both the L=143 Mpc and L=300 Mpc simulations) . Since these configurations correspond to different spatial resolutions, a smoothing was performed in the N=768$^3$ ionization boxes to roughly match the resolution of the 0.5 Mpc per cell achieved in the 300 Mpc simulation. Two important conclusions can be drawn from this figure: the excursion-set formalism allows to obtain a bubble power spectrum that matches the one obtained from a full N-body radiative transfer simulation when the volume size and cell resolution are the same. However, an important difference is observed between the power spectrum at large scales obtained in our simulation with a larger volume (L=300 Mpc) when compared to the lower volume ones. We believe this is due to the fact that we are finding larger bubbles in the 300 Mpc box, while smaller bubbles are absorbed in larger ones (a similar effect was observed in \citealt{mesinger07}). We have checked that such differences do not originate from the lower number of low mass halos obtained in the L=300 Mpc simulation 
or the fact that the mass function extends to larger masses in the larger simulation.
%%%%%%%%%%%%%%%%%

\subsection{The 21 cm signal}

With ionization maps at different redshifts we can focus on the predicted 21 cm signal from neutral hydrogen during the pre-EOR and the EOR. In the following Section we will briefly present the theoretical aspects behind this radiation and the results from our simulation. For a given observational frequency, the 21 cm signal corresponds to an intensity variation along the line of sight of
\begin{equation}
\centering
\delta T_b(\nu)=\frac{T_S - T_{\gamma}}{1+z}(1-e^{-\tau_{\nu_{21}}})\approx \frac{T_S - T_{\gamma}}{1+z}\tau_{\nu_{21}}\;,
\label{temp1}
\end{equation}
where $T_S$ is the spin temperature of the IGM, $z$ is the redshift corresponding to the frequency of observation $(1+z = \nu_{21}/\nu)$, where $\nu_{21}=$1420 MHz) and $T_{\gamma}$=2.73$(1+z)$K is the CMB temperature at redshift $z$. The optical depth corresponding to the hyperfine transition is given by \citep{field59}: 
\begin{equation}
\centering
\tau=\frac{3c^3\hbar A_{10}n_{HI}}{16k\nu_{21}T_S(1+z)(\partial V_r/\partial r)}
\label{tau}
\end{equation}
where $A_{10}$ is the spontaneous emission coefficient for the
transition ($2.85 \times 10^{-15}$ s$^{-1}$), $n_{\rm HI}$ is the
neutral hydrogen number density and $\partial V_r/\partial r$ is the comoving
gradient of the total radial velocity along the line of sight. Neglecting peculiar velocities, one has $\partial V_r/\partial r =
H(z)/(1+z)$.  The neutral density can be expressed as
$n_{\rm HI}=f_{\rm HI}X\rho_b/m_p$ where $f_{HI}=\rho_{\rm
  HI}/\rho_{\rm H}$ is the fraction of neutral hydrogen (mass
weighted), $X\approx 0.76$ is the hydrogen mass fraction, $\rho_b$ is
the baryon density and $m_p$ the proton mass. Since the baryonic content follows the dark matter distribution, the 21 cm temperature can then be written as: 
\begin{eqnarray}
\label{t21}
&&\delta T_b(\nu) \approx  23 f_{HI}(1+\delta)
\left(1-\frac{T_{\gamma}}{T_S}\right)
\left( \frac{h}{0.7}\right)^{-1}
\left( \frac{\Omega_b h^2}{0.02} \right)\times\nonumber\\ 
&&\left[\left(\frac{0.15}{\Omega_m h^2} \right)\left(\frac{1+z}{10}\right)
\right]^{1/2}\left(\frac{1}{1+1/H\,dv_r/dr}\right)\ \rm{mK},
\end{eqnarray}  
where $dv_r/dr$ is the comoving gradient of the line of sight component of the comoving velocity (obtained directly from the density field using a fourier transform). In high density regions of the simulation one can have $1/H\,dv_r/dr \leq -1$ and therefore we imposed a limit to this gradient of $dv_r/dr > -0.7 H$ in order to avoid singularities. Note that this only affects a very small number of cells ($<0.05\%$) and we verified that in neutral regions we have $|dv/dr| < H$.
In the above equation, both $f_{HI}$ and $\delta$ (as well as the velocity gradient) can be obtained as a direct result of our simulation (we used $f_{HI}=1-x_i$), 
but the term depending on $T_S$ is sensitive to variations of the spin temperature value over the volume. For $z < 10$, one can make the reasonable assumption that $T_s\gg T_{\gamma}$ since the IGM gas temperature should be well above 100K at these redshifts but at higher redshifts the contribution to the signal from the spin temperature has to be taken into account which we address further below. 

\begin{figure*}
\vspace{+0\baselineskip}
\centerline{
\includegraphics[width=0.48\textwidth]{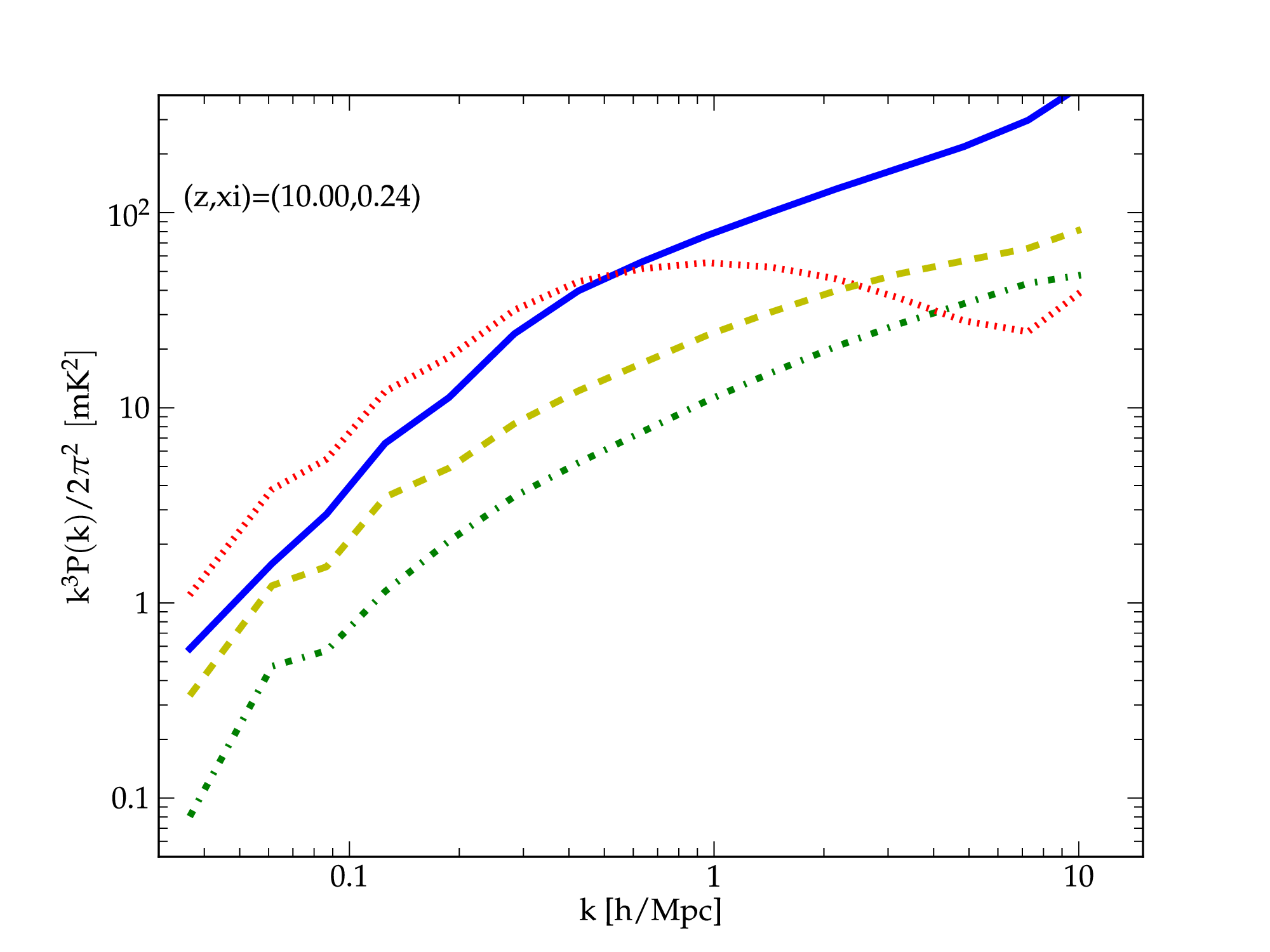}
%\hspace{-1cm}
\includegraphics[width=0.48\textwidth]{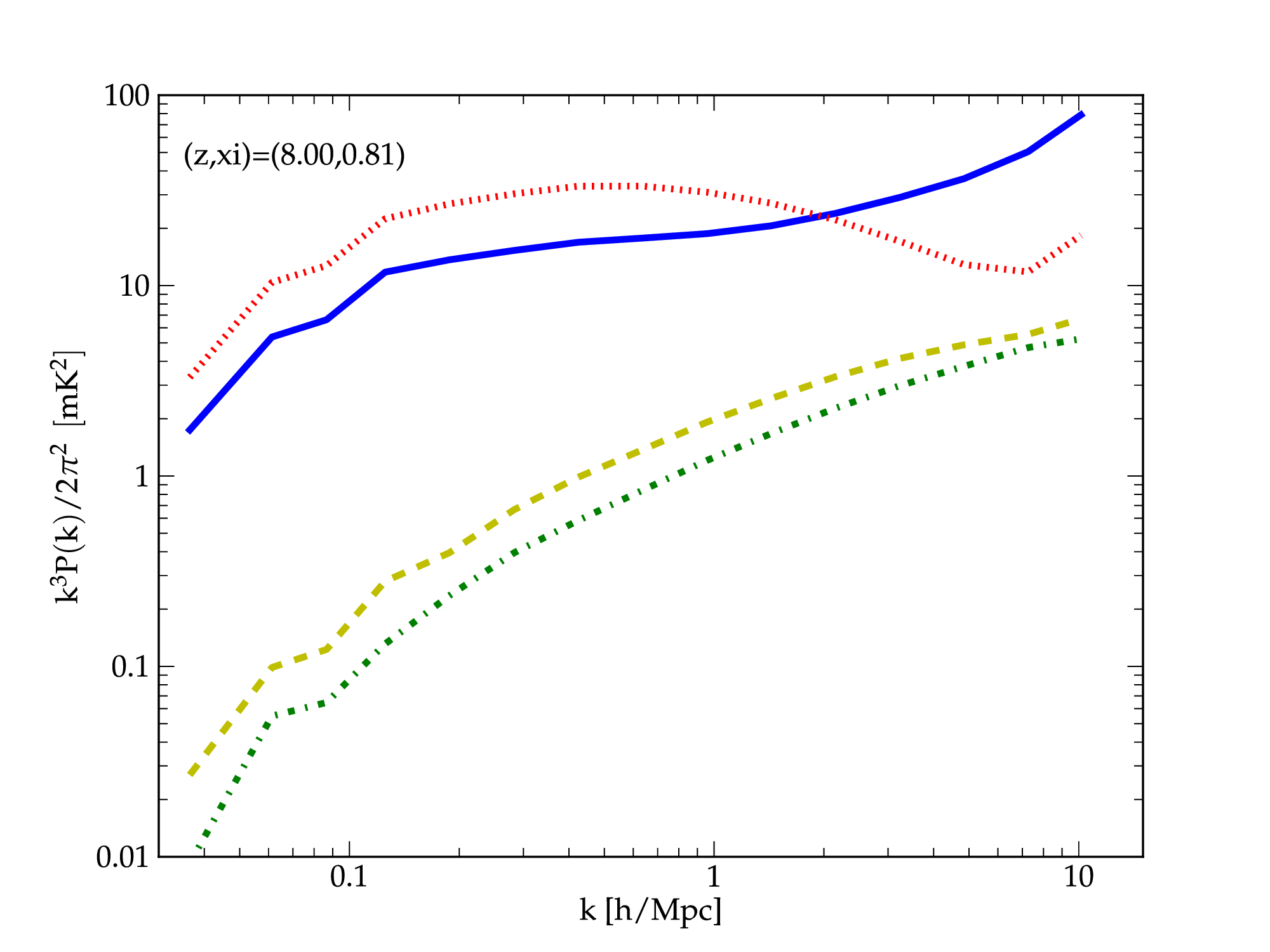}
}
\caption[]{
Brightness temperature power spectrum (solid blue line) and its different contributions at redshifts z=10 and z=8 with $\zeta=15.1$. The green dashed-dotted lines correspond to the power spectrum taking only velocity fluctuations, the yellow dashed lines only have density fluctuations and the red dotted lines only for ionization fluctuations.
\label{pk_temp}
}
\vspace{-1\baselineskip}
\end{figure*}

In figure~\ref{pk_temp} we plot the brightness temperature power spectrum for two different stages of the EoR ($\bar{x}_i=0.24$ and $\bar{x}_i=0.81$). In each figure we also plot the different contributions to the total power spectrum. This was done by using exclusively density, ionization  and velocity fluctuations at each time in eq.\ref{t21}. One can observe that at the initial stages of reionization (low $\bar{x}_i$), the main contributions to the 21 cm signal come from the density and ionization fluctuations, meaning that the hydrogen spatial distribution is moderately important due to the small ionization bubble size. At later stages, ionization bubbles have grown to a size where they completely dominate all the other contributions. As for the velocity fluctuations, they are sub-dominant at all stages, although being more important for higher redshifts.          
%%%%%%%%%%%%%%%%%%%%%
%%%%%%%%%%%%%%%%%%%5

\subsection{Spin temperature at high redshifts}

At high redshifts ($z\gtrsim 10$) the spin temperature ($T_S$) is no longer high enough to saturate the effect in equation \ref{t21} and we need to take into account the contribution of fluctuations from the spin temperature to the 21 cm signal. These originate from fluctuations in the coupling between the spin temperature and the gas temperature and the perturbations in the gas temperature itself ($T_K$) and we can write:
%%%%%%%
\be 
1-\frac{\tcmb}{T_S}=\frac{x_{tot}}{1+x_{tot}}\left(1-\frac{\tcmb}{ T_K}\right), 
\ee 
where $x_{tot}=x_\alpha+x_c$ is the sum of the radiative and collisional
coupling parameters. Collisions can be important for decoupling the HI 21 cm spin temperature from the CMB,
especially at very high redshifts ($z\gtrsim 30$) \citet{nusser05} and are straightforward to apply to the simulation \citep{allison69,zygelman05,furlanetto07,kuhlen06,hirata07}. The radiative coupling due to the absorption of $Ly_\alpha$ photons (the Wouthysen-Field effect, \citealt{wouthuysen52,field59}), on the other hand, should be dominant for $z<25$ and we shall concentrate on calculating this effect here. Note however that if we use a model where the Ly$_\alpha$ coupling is smaller, then the collisional coupling at lower redshifts can become important close to the sources due to the X-ray emission \citep{zaroubi07,kuhlen06}. The radiative coupling is given by
%%%%%%%%%%%%%%%%%%55
\begin{equation}\label{xa}
  x_\alpha=\frac{S_\alpha J_\alpha}{J_c},
\end{equation}
%%%%%%%%%%%%%%%%55
with $J_c \approx 5.552 \times 10^{-8}(1+z)\ {\rm m^{-2} s^{-1} Hz^{-1} sr^{-1}}$
and $S_\alpha$ is a correction factor of order unity \citep{chen04,hirata06,
 chuzhoy07,furlanetto06d,furlanetto06c}. We follow the prescription in \citet{santos08} to calculate $J_\alpha$ (the spherical average of the number of
\Lyman photons hitting a gas element per unit proper area per unit time per
unit frequency per steradian), given by a sum over the hydrogen
levels $n$, 
\bea
\label{Ja} 
J_{\alpha}(\bx,z) \ &=&\ \frac{(1+z)^2} {4 \pi}
\sum_{n=2}^{n_{\rm max}} f_{\rm rec}(n)\times\\ \nonumber
&\times& \int {d\Omega'\over 4
  \pi}\int_0^{x_{\rm max}(n)} dx'\ \psi(\bx+\bx',z')\ \epsilon_\alpha(\nu')\,
\eea
where we used $n_{\rm max}=16$, $f_{\rm rec}(n)$ is the fraction of Lyman-n photon that
cascade through \Lyman and the redshift $z'$ in equation
\ref{Ja} is such that $x'=\int_z^{z'} cH^{-1} dz''$. 
The emitted frequency at $z'$ is given by \be \nu_n' = \nu_{\rm LL} (1-n^{-2}) {(1+z')\over (1+z)}\ ,
\label{nu} \ee in terms of the Lyman limit frequency $\nu_{\rm LL}$
and $x_{\rm max}(n)$ corresponds to the comoving distance between $z$
and $z_{\rm max}(n)$ such that: \be 1+z_{\rm max}(n) = (1+z)
{\left[1-(n+1)^{-2}\right] \over(1-n^{-2})} \ .  \ee 

\begin{figure}
\includegraphics[scale=0.44]{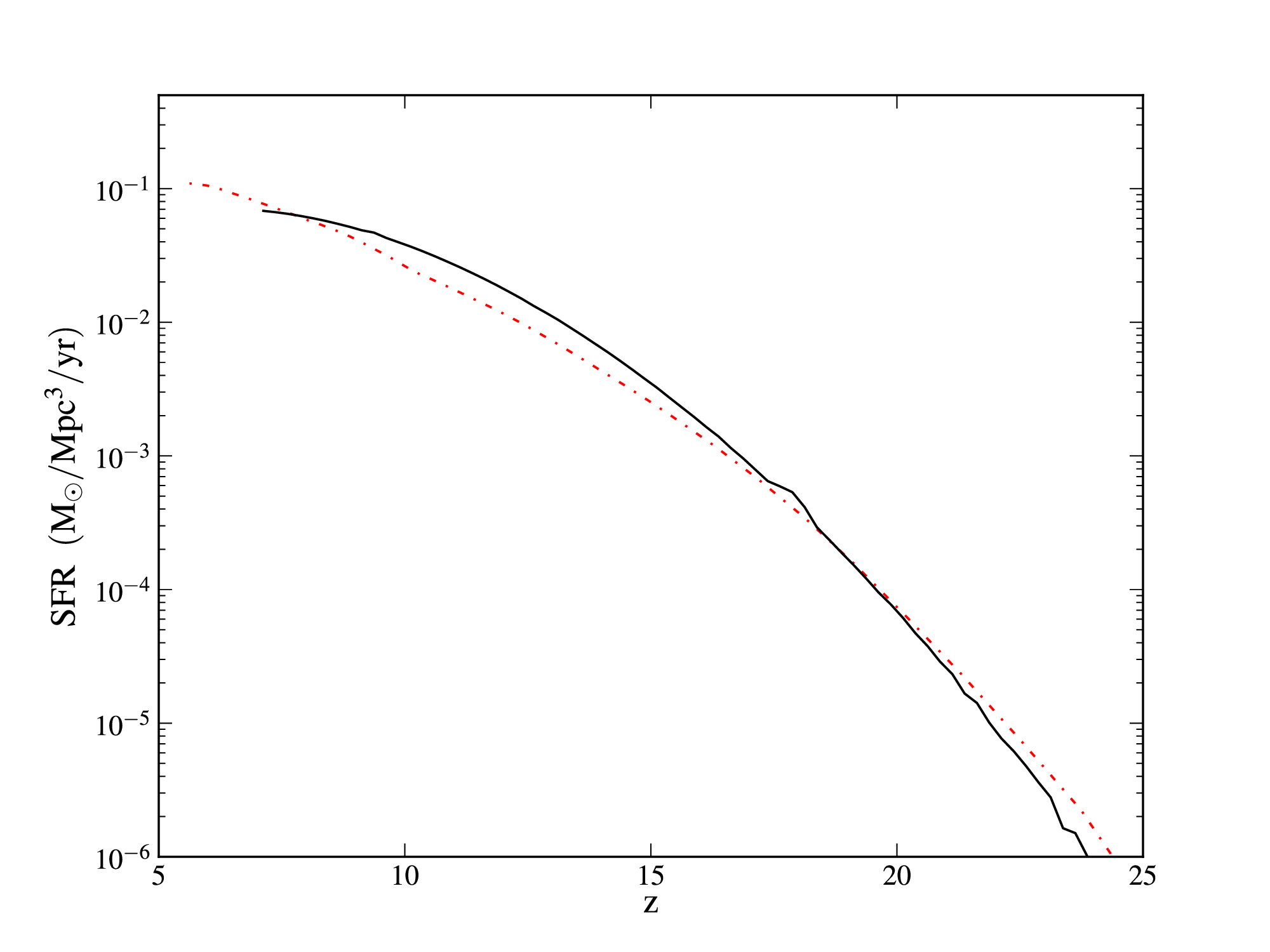}
\caption{Average comoving star formation rate as a function of redshift. Black solid line is from our simulation and red dashed-dotted line is from an N-body simulation in \citet{shin07}.}
%\vspace{0.5cm}
\label{sfrd}
\end{figure}

In equation~\ref{Ja}, $\psi(\bx,z)$ is the comoving star formation rate density and $\epsilon_\alpha(\nu)$ gives the spectral distribution function of the
sources (defined as the number of $Ly_n$ photons per unit frequency emitted
at $\nu$ per baryon in stars). We can easily assume any model for $\epsilon_\alpha(\nu)$ in our code to test for different sources of radiation. For this simulation we used: $A\nu^{-\alpha}$, with $\alpha=0.9$ and $A$ set such that we get a total of 20000 \Lyman photons per baryon which gives a $Ly_n$ emission similar to the one used in \citet{santos08}. Note however that the amplitude is degenerate with the normalization of the star formation rate.
In order to obtain the star formation rate density from the simulation we used the halo catalogues generated to create boxes of collapsed mass ($M_{coll}(\bx,z)$) for two closely separated redshifts and calculated: $\psi(\bx,z)=f_*{d\over dt}M_{coll}(\bx,z)$, where $f_*$ was used to normalize the star formation rate density 
to the one used in \citet{shin07} (see fig. \ref{sfrd}). 
%%%%%%%%%%%%
%%%%%%%%%%%%%%% 

\begin{figure*}
\centerline{\includegraphics[scale=0.43]{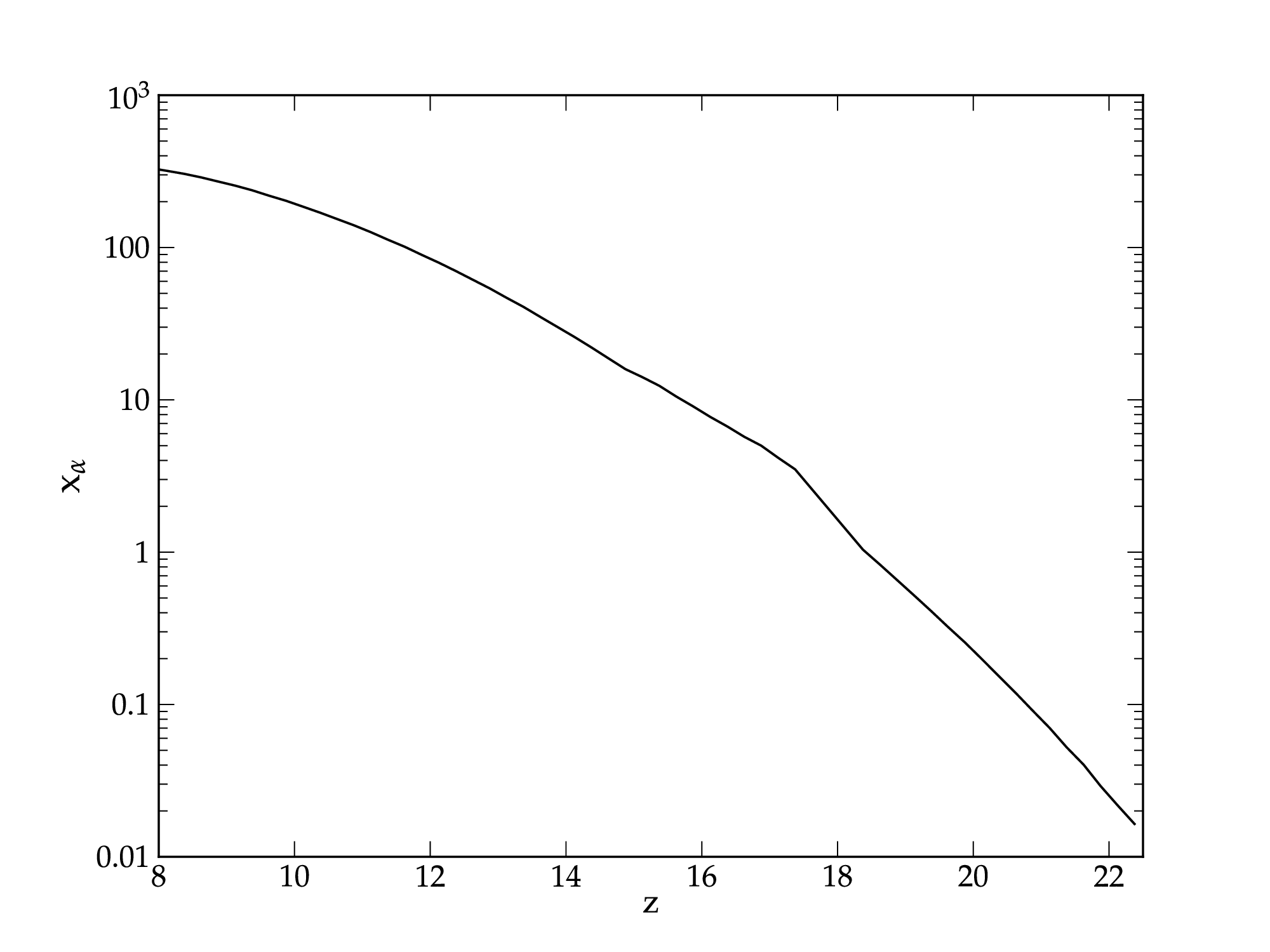}  \includegraphics[scale=0.43]{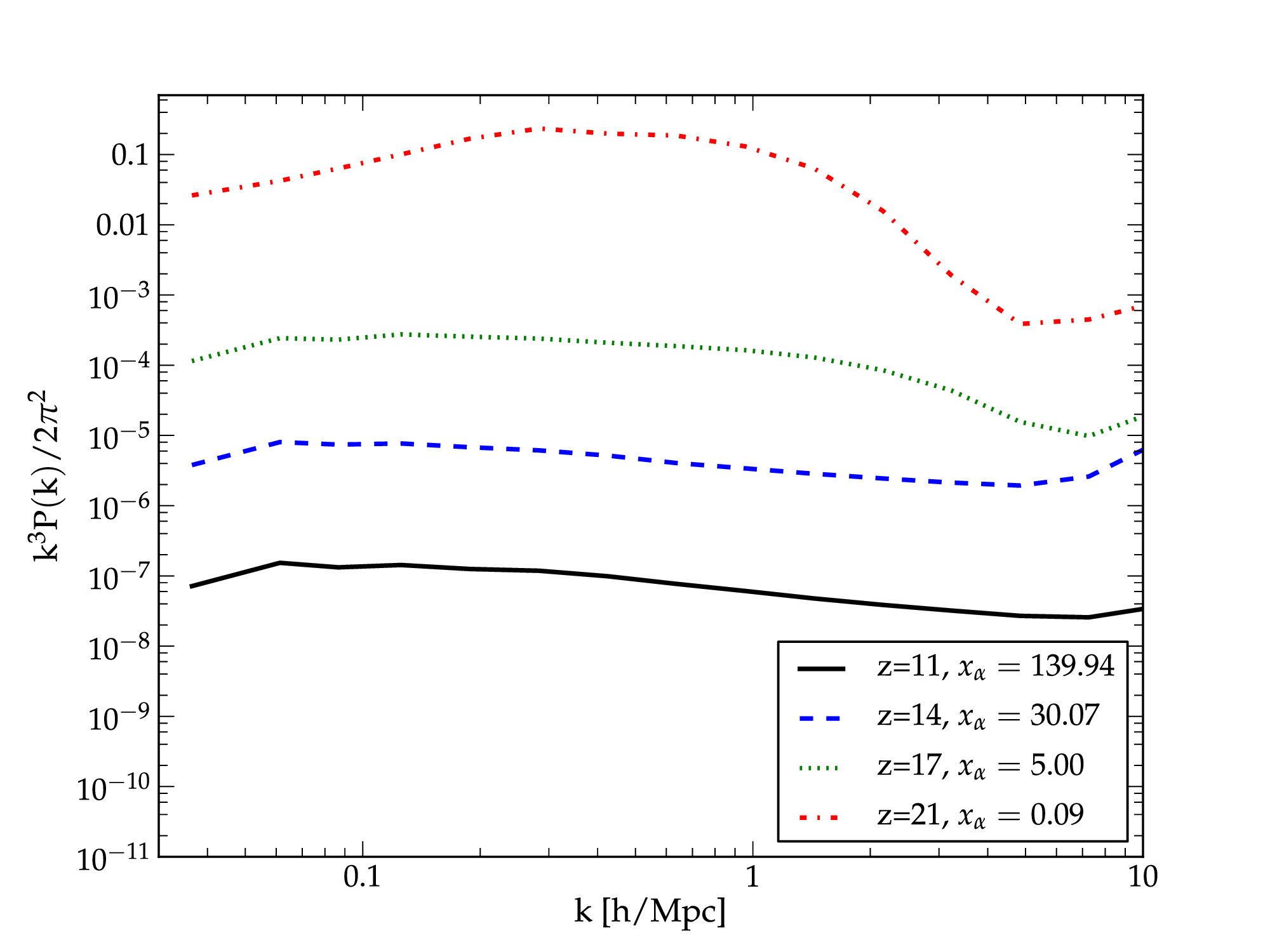}}
\caption{Left: Evolution of the $<x_\alpha>$ with redshift.
Right: power spectra of $x_\alpha/(1+x_\alpha)$ divided by the square of the average of this quantity.}
\vspace{0.5cm}
\label{fig_xa}
\end{figure*}

Finally, to make the code faster and taking into account that we only have the star formation rate density boxes for a range of redshifts, we divided the integral over $d\Omega'\ dx'$ above in eq. \ref{Ja} into redshift bins, e.g.
\be
\sum_{i=0}^{i=imax(n)}\int d^3x' {\rm K}_i(x')\ \psi_i(\bx+\bx'),
\ee
with kernel ${\rm K}_i(x')=\epsilon_\alpha(\nu')/(4\pi\ {x'}^2)$ if $z_i \leq z' < z_{i+1}$ corresponding to the redshift bin for which we assume the $\psi_i$ box to be constant and ${\rm K}_i(x')=0$ otherwise. Here, $imax(n)$ is set by $x_{max}(n)$ but to avoid having to sum over too many boxes we made the assumption that when $x'>71$ Mpc the star formation rate was homogeneous. We have checked by changing this limit that it is a good approximation (note that the star formation rate density fluctuations are quite small on large scales, $k<0.1$ Mpc$^{-1}$). Moreover, since this is a convolution, the expression above can be easily calculated with resort to a Fast Fourier Transform (FFT):
\be
\sum_{i=0}^{i=imax}FFT\left(FFT[{\rm K_i}]\cdot FFT[\psi_i]\right).
\ee
Figure~\ref{fig_xa}, left panel, shows the evolution of the radiative coupling with redshift which is already larger than 1 for $z<18$. The right panel on the other hand shows the power spectrum of the quantity $x_\alpha/(1+x_\alpha)$ which is what contributes to the 21 cm signal (we divided the power spectra by the square of the average for later comparison, although this does not change the amplitude much). We see here that the fluctuations peak around $z=20$ when $x_\alpha\approx 0.1$.
%%%%%%%%%%%%
%%%%%%%%%%%%%%% 

\begin{figure*}
\centerline{\includegraphics[scale=0.43]{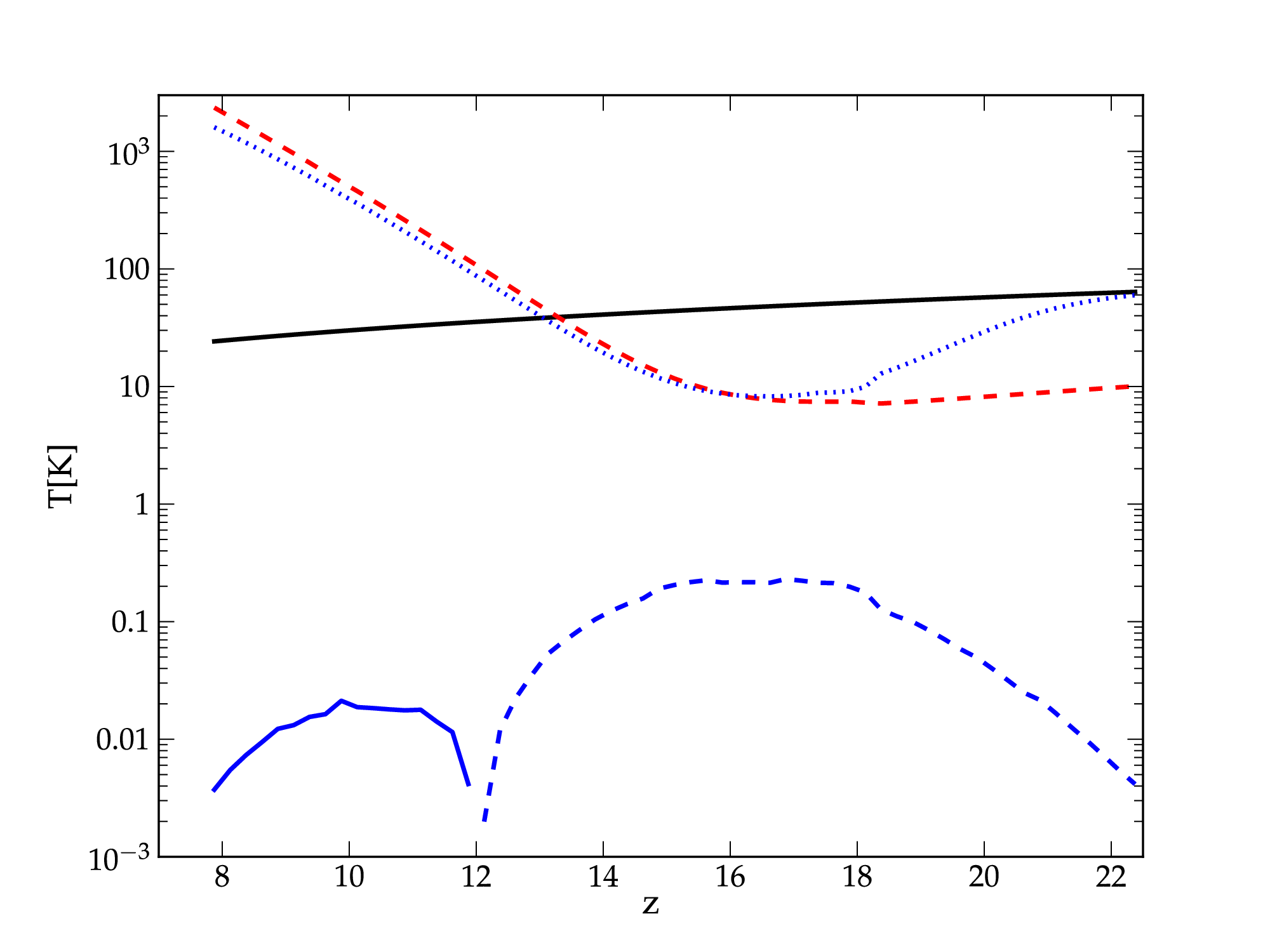}  \includegraphics[scale=0.43]{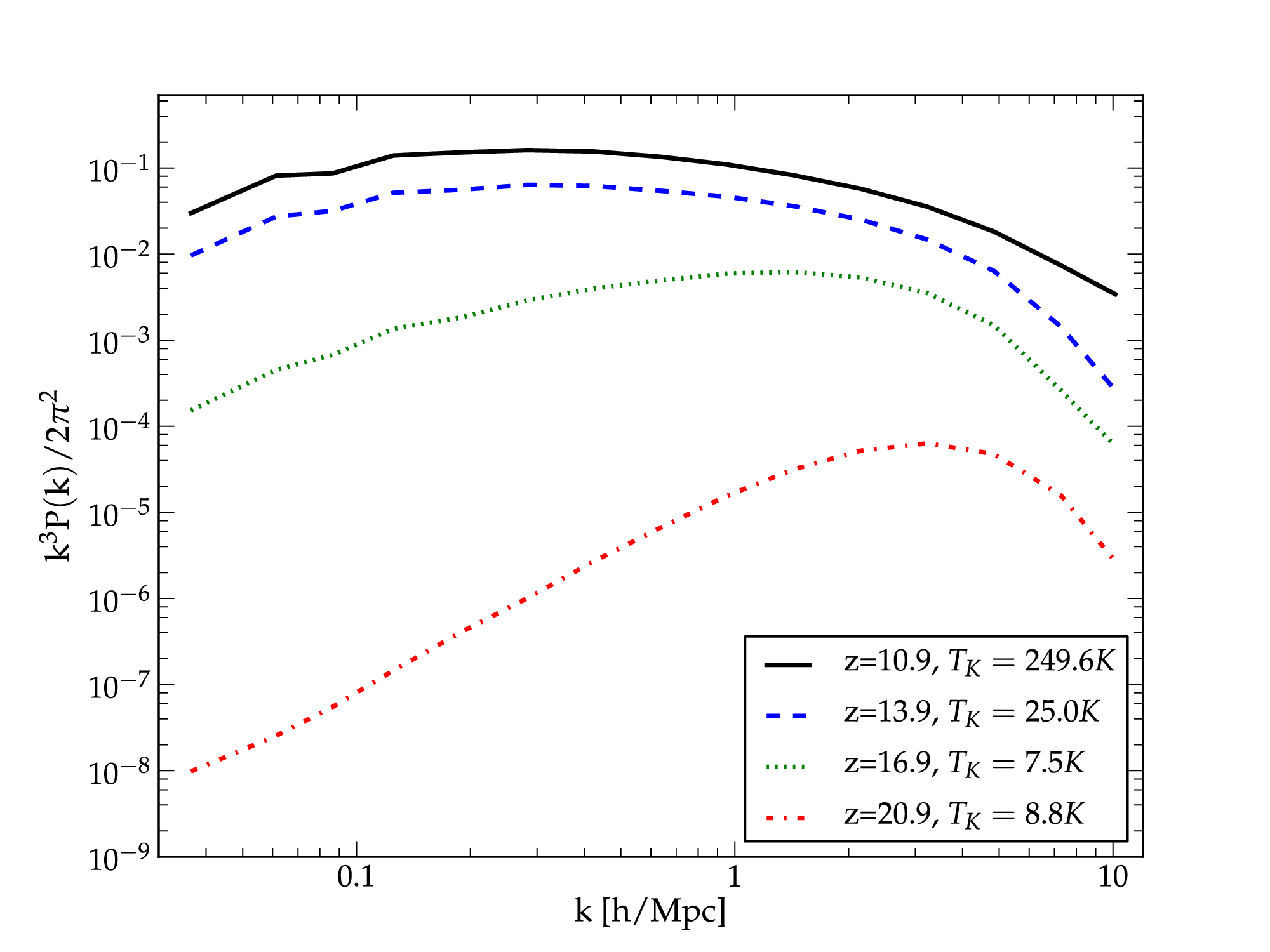}}
\caption{Left: The evolution of the gas temperature, spin temperature and brightness temperature with redshift. Solid black line - CMB; Red dashed - gas temperature; Blue dotted - spin temperature; Dark blue solid/dashed (below) - brightness temperature. 
Right: power spectra of the factor $1-T_{CMB}/T_K$ divided by the square of its average.}
%\vspace{0.5cm}
\label{temp}
\end{figure*}

The gas in the IGM is also heated as the reionization progresses. In our simulation, we assume that this heating is done by x-rays with an emission connected to the star formation rate (see \citealt{santos08}). Our starting point was to calculate the total x-ray energy per unit time deposited in a given cell:
\bea
\epsilon_{Xn}(\bx,z)=(1+z)^2\sum_i {n_i\over n}\times\nonumber\\ 
\int d^3x' \psi(\bx+\bx',z')\ {\rm K}(x'),
\eea
where
\be
{\rm K}(x')=\int_{\nu_{\rm th}}^\infty d\nu
{\sigma_i(\nu)\over 4\pi |\bx'|^2} {\epsilon}_X(\nu)e^{-\tau(z,x')}(h\nu-h\nu^i_{\rm th})
\ee 
and $n_i$ ($i=$HI, HeI, HeII) is the number density (due to our prescription for ionization, we assumed that $f_{HeI}=f_{HI}$ and there was no HeIII at these high redshifts). The optical depth $\tau$ should in fact depend on the path between $\bx'$ and $\bx$ but to increase the speed of the calculation we assumed $n_i$ to be homogeneous in the integral used to obtain the optical depth. Moreover, we tabulated the kernel values (which only depend on $x'$) and interpolated them when doing the FFT making the code faster by a large factor. In the equation above, ${\epsilon}_X(\nu)$ is again assumed to be a power law, with a spectral index and amplitude compatible with what is observed for starburst galaxies (note that this amplitude is also degenerate with the star formation rate normalization). We then calculate how much of this energy is used for secondary ionizations in the neutral IGM,
%%%%%%%%%%%%%%%%%%%%%%%%%
\begin{equation}\label{xehistory}
\frac{\ud x_e}{\ud t}=\epsilon_{Xn} {f_{\rm ion}\over E_{\rm th}}
\end{equation}
%%%%%%%%%%%%%%%%%%%
as well as how much is used to heat the IGM, by integrating:
\begin{equation}\label{thistory}
  \frac{\ud T_K}{\ud t}=\frac{2T_K}{3n}\frac{\ud n}{\ud t}+\frac{2\epsilon_{Xn} f_{\rm heat}}{3 k_B},
\end{equation}
where $f_{\rm ion}$ is the fraction of energy converted into ionizations and $f_{\rm heat}$ the fraction converted into heat \citep{shull85} which depends on the fraction of free electrons, $x_e$. We started the integration at $z=25$ assuming an initial temperature of $T=14$K consistent with adiabatic cooling of the gas from $z=150$ (where it is approximately equal to the CMB temperature).
Figure~\ref{temp}, left panel, shows the evolution of the gas temperature with redshift, where we can see that most of the IGM is heated above 100K for $z\lesssim 11$. The spin temperature starts very close to the CMB one (no signal) and approaches the gas temperature as $x_\alpha$ increases, following it for $z<17$.
In the right panel we show the power spectrum of the quantity $1- T_{CMB}/T_K$ which contributes directly to the fluctuations in the brightness temperature (but in this case we divided by the square of the average). Comparing to the right panel in fig. \ref{fig_xa} we see that fluctuation in $x_\alpha$ are dominant at very high redshifts ($z>18$ when $x_\alpha < 1$) while the perturbations in the gas temperature are more important at lower redshifts, peaking at $z=14$ if we multiply the fluctuations by the average of $1- T_{CMB}/T_K$ which is what shows up in the 21 cm signal.
%%%%%%%%%%%%
%%%%%%%%%%%%%%% 

Finally, putting it all together we can calculate the 21 cm signal using equation \ref{t21}. Calculation of $x_\alpha$ and the gas temperature takes around 5 minutes for each redshift in our machine using 20 CPUs. This obviously depends on how many star formation rate density boxes we use for the integration which in turn depends on the assumed redshift bin (in our case we used $dz=0.25$ for each box).
In figure~\ref{temp}, left panel, we can also see the evolution of the average brightness temperature with redshift, which is observed in absorption down to $z=12$ where it becomes approximately zero when $T_K\approx T_{CMB}$. 

\begin{figure*}
\centerline{\includegraphics[scale=0.45]{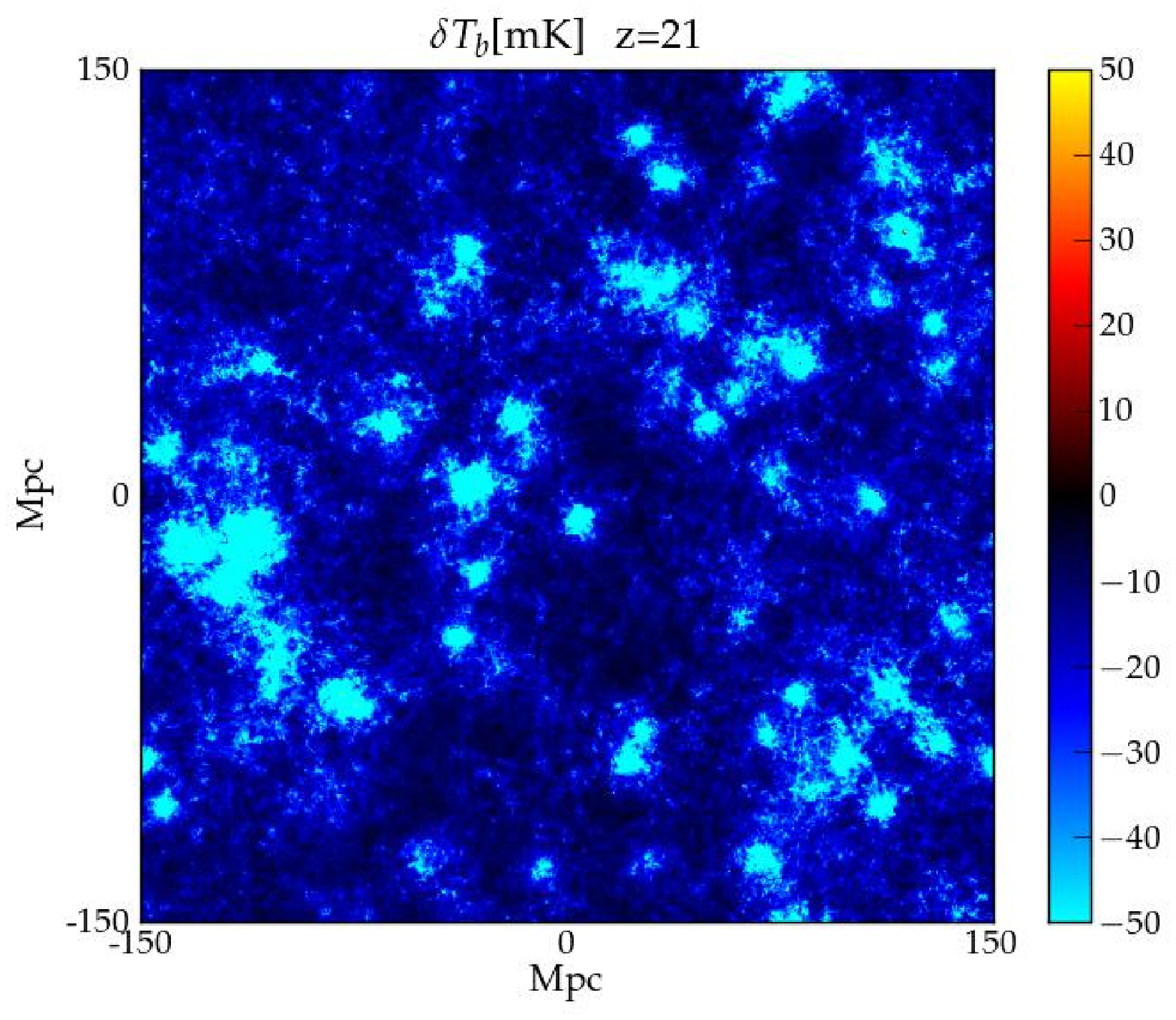}  \includegraphics[scale=0.45]{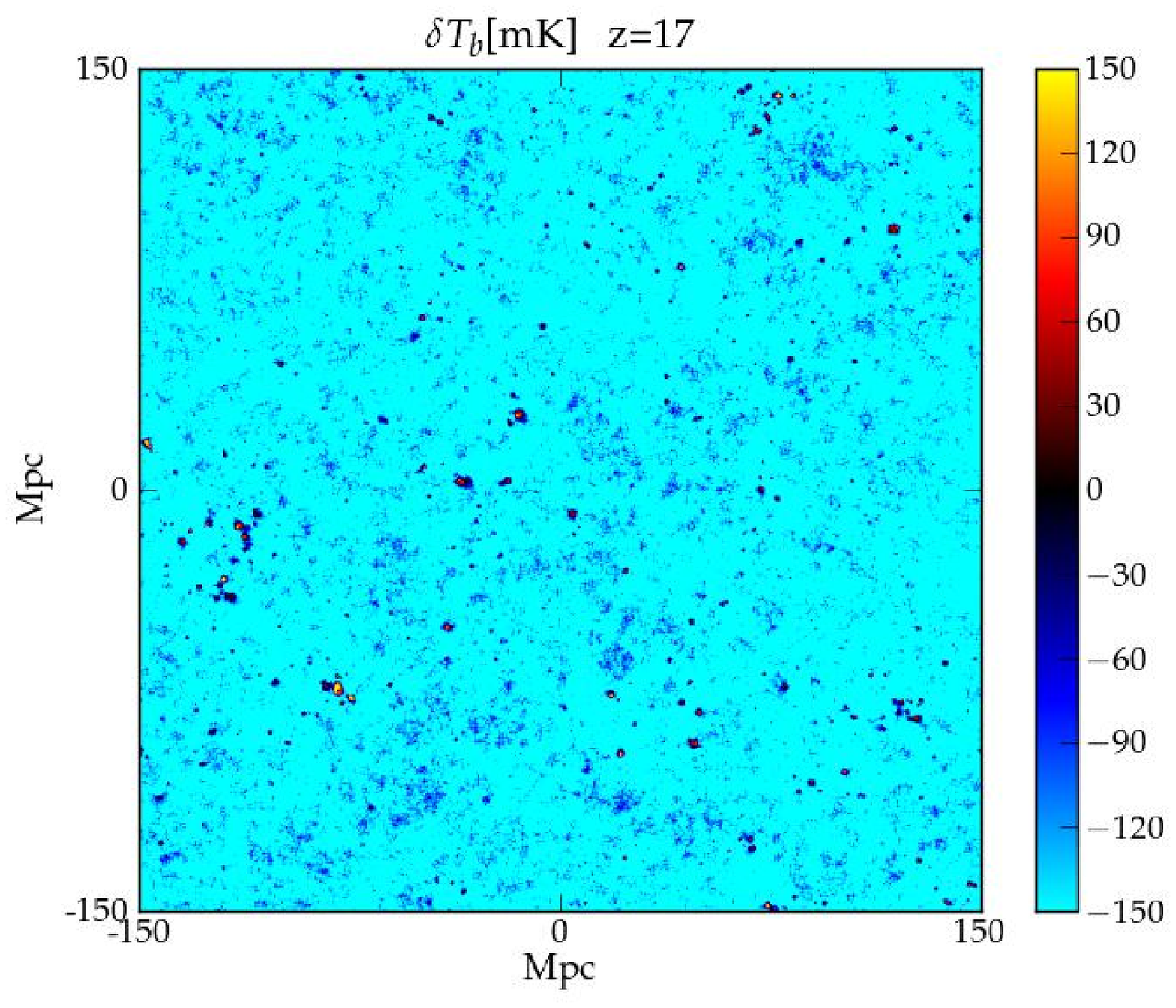}}
\centerline{\includegraphics[scale=0.45]{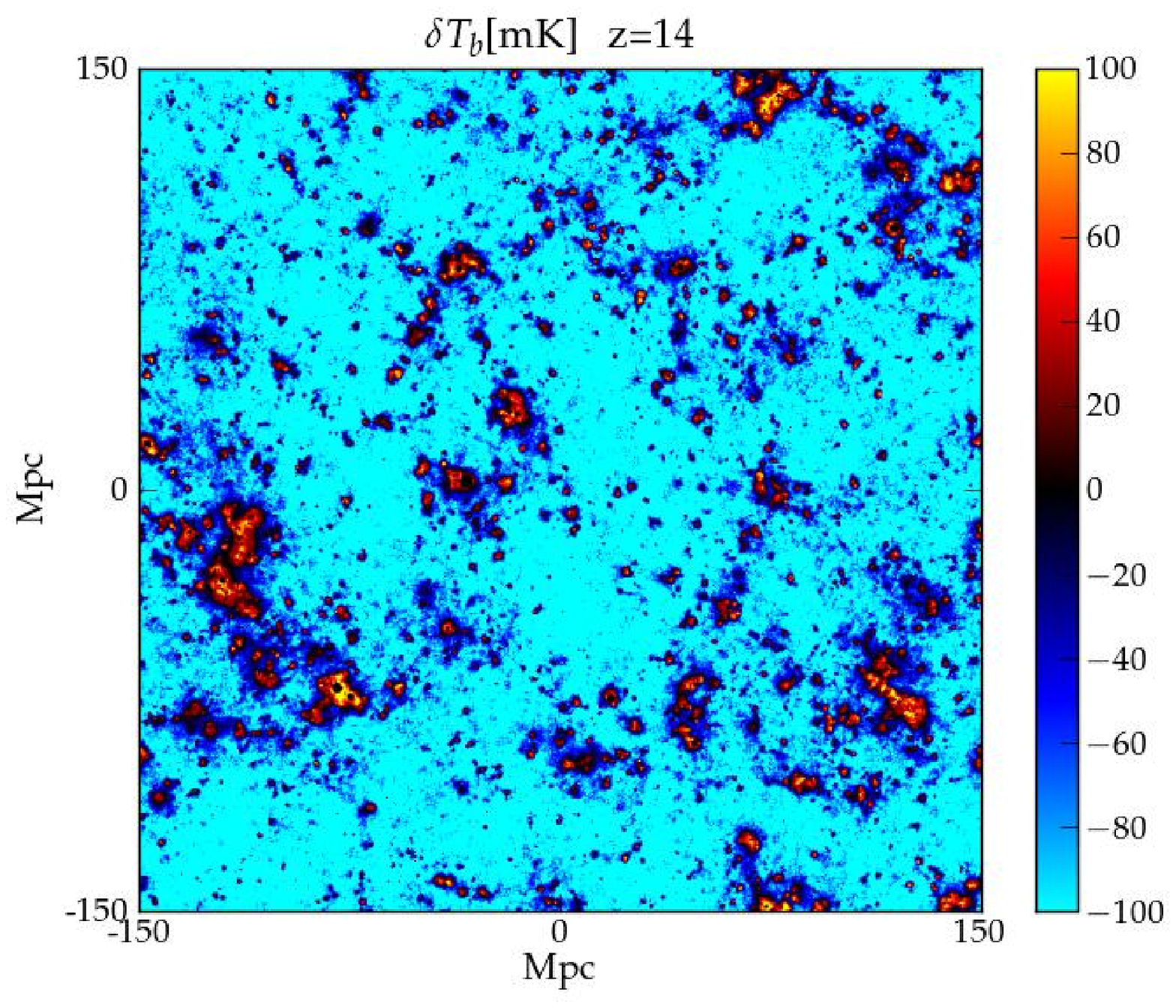}  \includegraphics[scale=0.45]{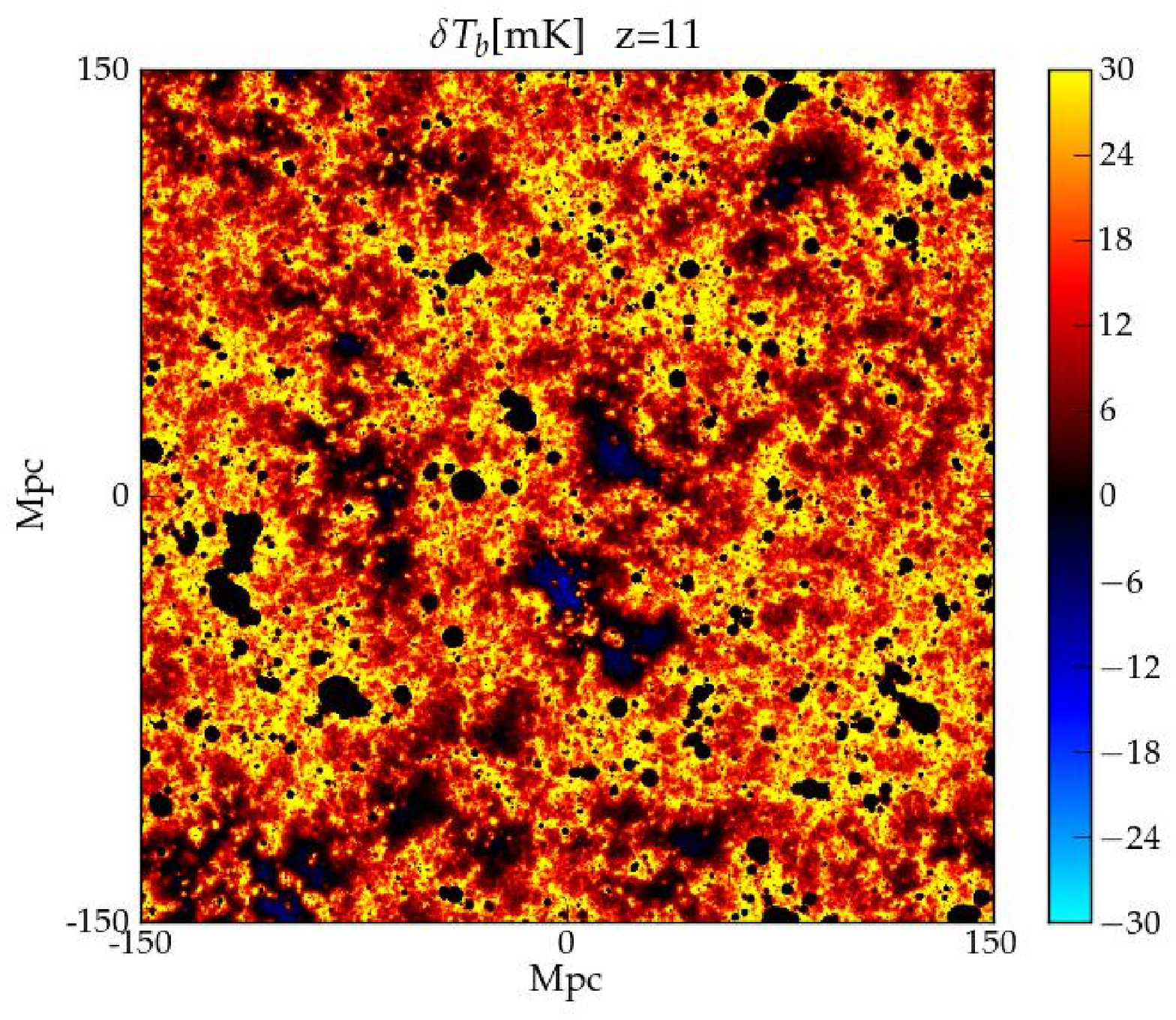}}
\caption{Maps of the 21 cm brightness temperature at very high redshifts from the simulation,  including all sources responsible for brightness temperature fluctuations.}
\label{map_21}
\end{figure*}

\begin{figure}
\includegraphics[scale=0.44]{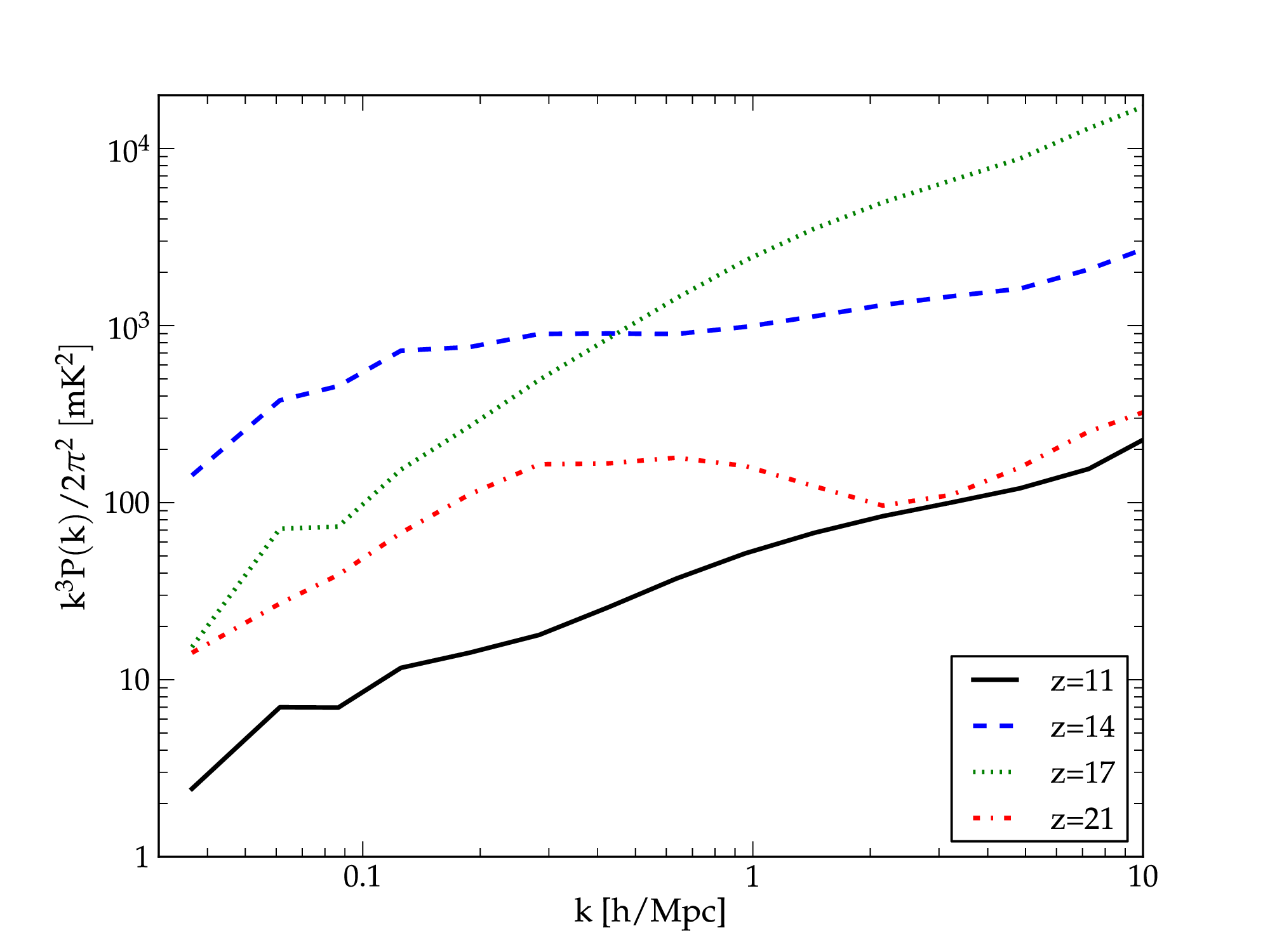}
\caption{The power spectrum of the 21-cm signal at high redshifts including all fluctuations in the spin temperature.}
\label{ps_21}
\end{figure}

In figure~\ref{map_21} we show maps of the 21 cm signal for a few very high redshifts taking into account all the effects.
At high redshifts $(z=21)$ most of the temperature is zero ($T_s\sim T_{CMB}$) but we already start to see some cold spots where the \Lyman sources couple the spin temperature to the cold gas temperature. At $z=17$ the spin temperature is basically following the gas temperature everywhere and we can see fluctuations in the 21 cm signal due essentially to the gas temperature (although most of the Universe is still cold at this epoch). At $z=14$ we can already see the higher temperature regions surrounding the light sources located in the halos. Finally at $z=11$ basically all the IGM is heated above the CMB and we start seeing the ionization bubbles ($\bar{x}_i\approx 0.1$, but note that some of the dark regions here are not due to the ionized bubbles but because $T_S=T_K\sim T_{CMB}$).
Figure~\ref{ps_21} shows the corresponding power spectrum.
Note again that at $z\approx 11$ the fluctuations in the gas temperature still make an important contribution to the 21 cm signal, especially on large scales (increasing the power), even though the IGM temperature is already above the CMB one ($T_K\sim 150$K and $T_{CMB}=32.7$K). 
%%%%%%%%%%%%
%%%%%%%%%%%%%%%
%%%%%%%%%%%%
%%%%%%%%%%%%%%%

\section{Extending the simulation to very large volumes}

Although quite fast and based on first principle physics, the semi-numerical simulation presented so far still imposes some limitations regarding the volume it can cover. In the absence of large computing resources, it is not feasible to run a simulation covering a large dynamical range. For example the simulation presented above, covering 300$^3$ Mpc$^3$, required to push the limits of our machine, using 20-processors and almost 64 GB of shared RAM in order to perform the FFT filtering. The main problem with the algorithm that was presented is that it still requires a very large number of cells in order to achieve sufficient resolution to filter scales corresponding to 10$^8$M$_{\odot}$ halos and cover large volumes at the same time. If we want to simulate the very large volumes that will be surveyed by the next generation radio telescopes a less demanding algorithm is required. 

This motivated us to implement a slightly simpler method when identifying the halos. Using the same resolution of N=1800$^3$, but with L=1000 Mpc (which corresponds to roughly a 5 $\times$ 5 deg$^2$ sky at z=15) the minimum cell size of $L/N\sim$0.51 Mpc corresponds to a minimum halo mass of $\sim$ $1\times 10^{10}$ M$_{\odot}$. Clearly, this configuration does not allow us to resolve halos down to the required minimum mass of 10$^8$ M$_{\odot}$. However, one can use the expected halo mass function to place these smaller halos in our large volume simulation cells by using the prescription of \citet{wilman08}. It follows directly from the definition of the mass function that the quantity $n=dn/dM \Delta M \Delta V$ corresponds to the mean number of halos found in a given cell with comoving volume $\Delta$V. If one would simply perform a Poisson sampling for each cell labelled (i,j,k), with mean n(i,j,k), the distribution of obtained halos would yield a correct mass function but their positions would be completely random. In order to correlate the halo positions with the underlying density field, a normalized bias term can be added,
\begin{equation}
\centering
n=K e^{b(z,M)\delta(z)}\frac{dn}{dM}\Delta M\Delta V\;,
\label{poisson1}
\end{equation}
where $b(z,M)$ is the bias model for halos at a given redshift and $\delta(z,M)$ the density field at the same redshift. The constant K is a normalization constant that ensures the consistency of the above expression with the mass function $\frac{dn}{dM}$ when averaging over the large volume box. For a Gaussian field, this requirement leads to:
\begin{equation}
\centering
K=e^{-b(z,M)^2 \sigma(z)^2 /2}.
\label{poisson2}
\end{equation}
Here, $\sigma(z)^2$ is the variance of the density field at a typical scale of cell size. The above equations introduce a dependence of the halo locations on the density field, by amplifying the overdense regions with respect to underdense ones. As for the bias term, it describes the difference in clustering between dark matter halos and the mass density field. We consider a linear bias developed from the Press-Schechter formalism by \citet{mo96} and later modified by \citet{Jing98}:
\begin{equation}
\centering
b(z,M)=\left(1+\frac{\nu-1}{\delta_c}\right)\left(\frac{1}{2\nu^4-1}+1\right)^{0.06-0.02n}\;,
\label{bias}
\end{equation}
where $\nu=\delta_c/\sigma(M)$ and n is the spectral index of the primordial density field fluctuation power spectrum.

\begin{figure}
\includegraphics[width=0.45\textwidth]{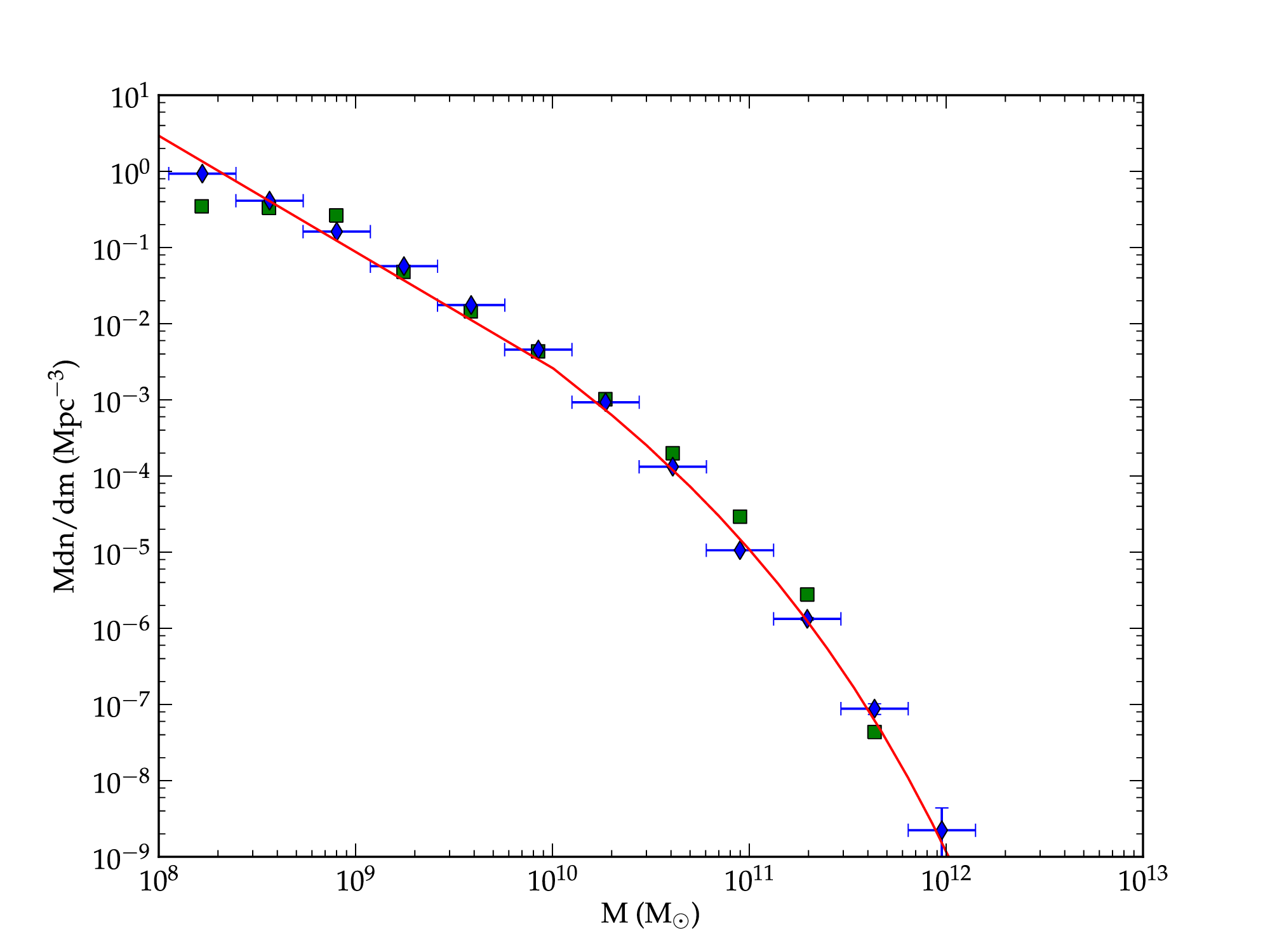}
\caption[]{Mass function at z=10 for our simulation with L=300Mpc and N=1800$^3$ (green squares) comparing with the large volume simulation with L=1000 Mpc and N=1800$^3$ (blue diamonds). Halos with masses M $<$ 10$^{10}$M$_{\odot}$ were placed using the Poisson sampling method introduced in this section. }
\label{mass_function_poisson}
\end{figure}

We have then used this prescription in complement with the excursion-set formalism to perform a dark matter simulation with L=1000 Mpc. Since the cell size defines the lower mass scale using the exclusion set formalism, halos with mass above 10$^{10}$ M$_{\odot}$ were obtained using this method. For scales bellow the cell size, we did a Poisson sampling of the entire box using eq. \ref{poisson1} for the distribution mean. In this way, we allow for the same cell to contain more than one halo, as long as the sum of their associated volumes does not exceed the cell volume $\Delta V$. In figure~\ref{mass_function_poisson}, we plot (blue diamonds) the obtained mass function for this run and compare it to the higher resolution, L=300 Mpc simulation obtained using only the excursion set formalism. We can clearly see that only this method allows to have a high compatibility with the expected mass function with low resolution, whereas the full excursion-set formalism only has similar results when the cell resolution is considerably high. 
%%%%%%%%%%%%%%%%%%5
%%%%%%%%%%%%%%%

\begin{figure}
\includegraphics[width=0.45\textwidth]{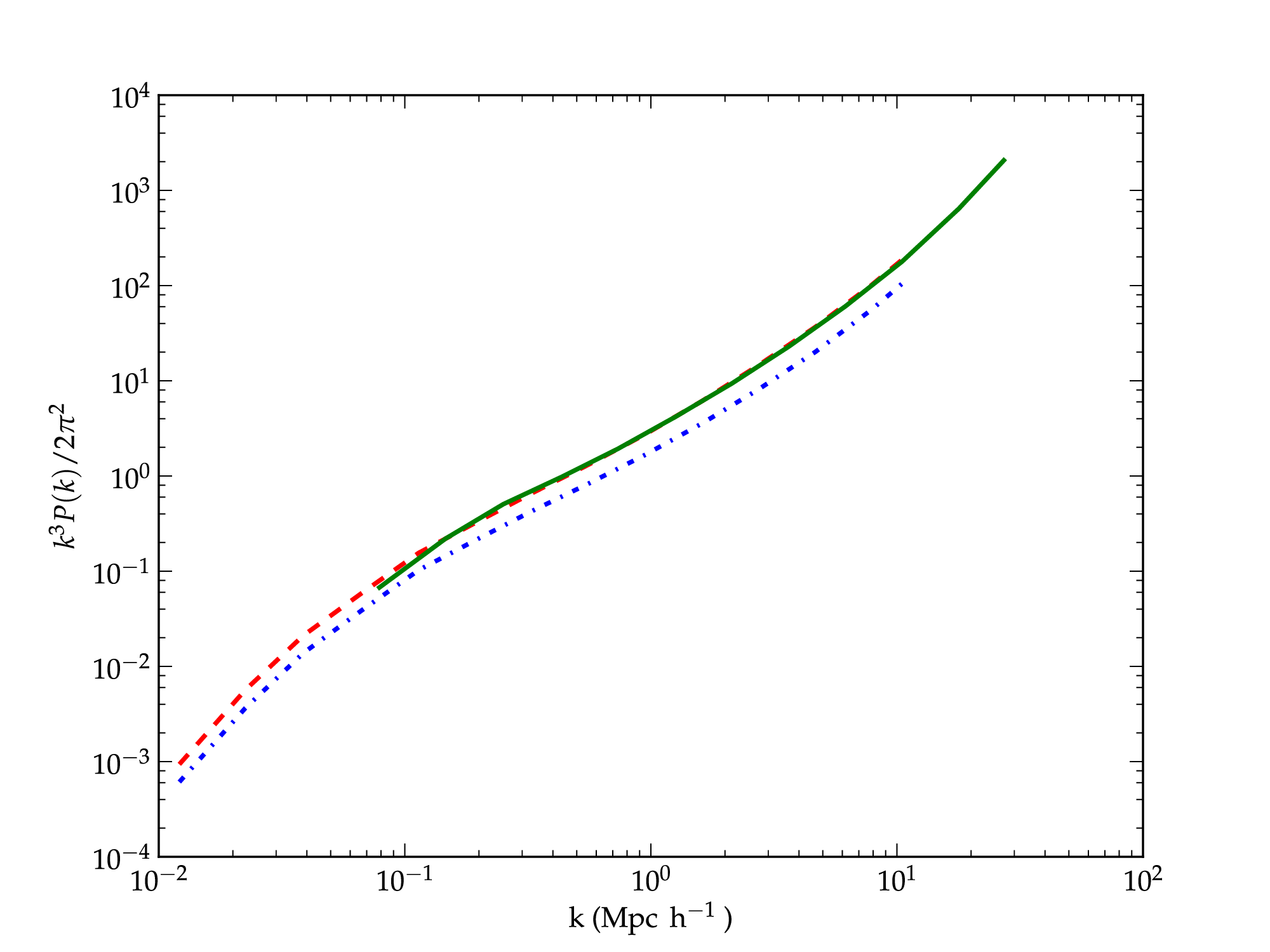}
\caption[]{Halo power spectrum of the halo field obtained in our very large volume simulation before and after applying the Zel'dovich corrections (dotted dashed  and dashed curves, respectively), compared with the result from the L=143 Mpc of \citet{trac08}.}
\label{deltahh_poisson}
\end{figure}

Despite the good agreement with the mass function, a more important aspect is to see how well this formalism can reproduce the spatial distribution of the halos and correlation with the density field. This is a crucial factor, since a correct determination of the reionization bubbles depends on the relation between these quantities. One way of testing for this is by computing the halo power spectrum, P(k)$_{hh}$, of the quantity $\delta_{hh}=M_{coll}/\bar{M}_{coll}-1$. In figure \ref{deltahh_poisson} we show the obtained power spectrum, and compare it with the one from the simulation of \citet{trac08}. The halo power spectrum derived from our method shows the same characteristic shape as the one obtained with the full N-body simulation. Once we introduce the first order Zel'dovich corrections for the halo positions (red dashed curve) the power spectrum becomes almost in perfect agreement with the simulation (we also checked that this agreement is true for different mass bins). Such result shows that this somehow ad hoc method can retrieve the ``correct'' statistical properties of dark matter halos.
%%%%%%%%%%%%%%%%%%%%%%%%%%%%%%%%

\begin{figure*}
\centerline{\hspace{-0.1cm}\includegraphics[scale=0.43]{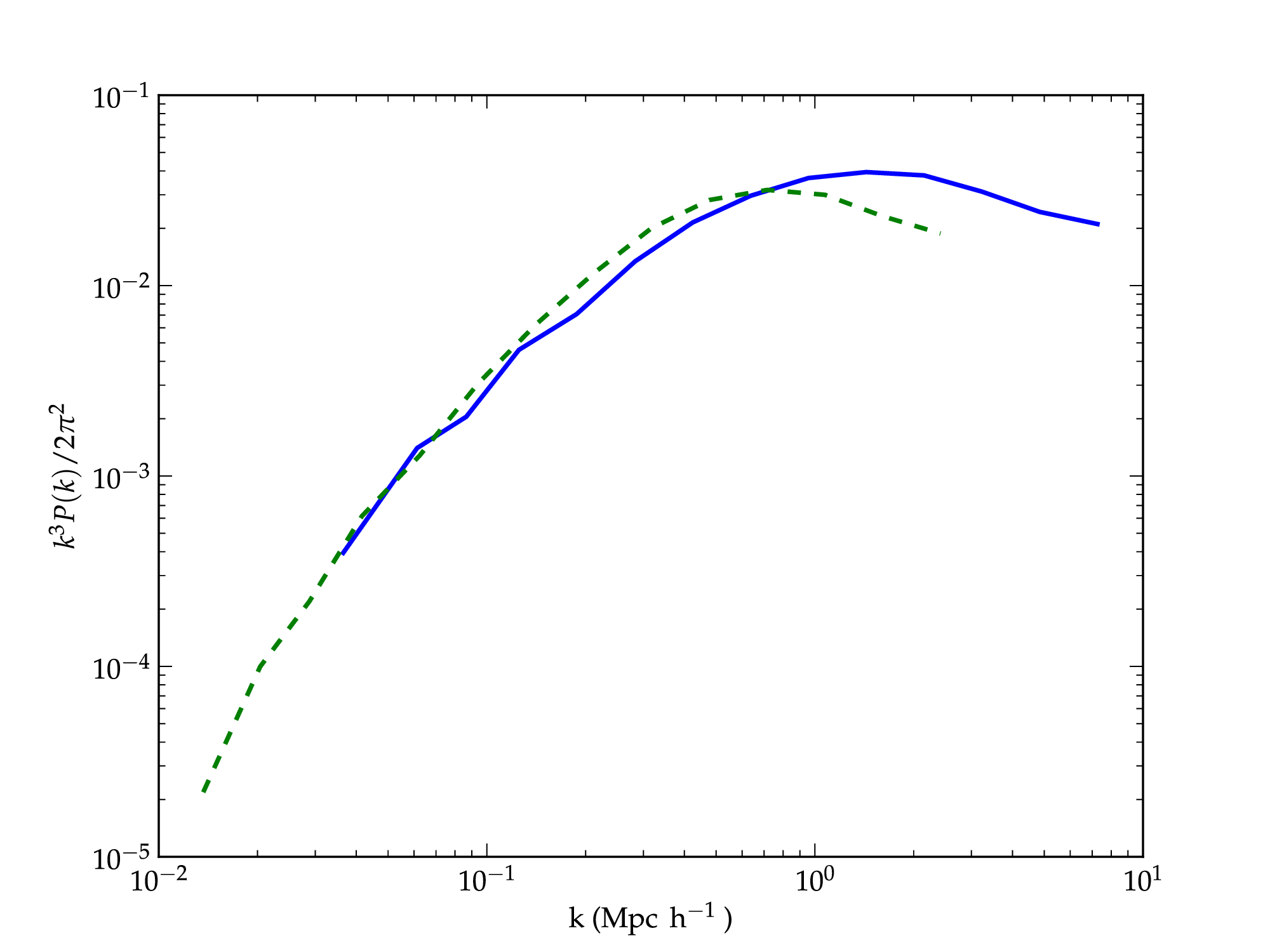}\hspace{0.0cm}  \includegraphics[scale=0.43]{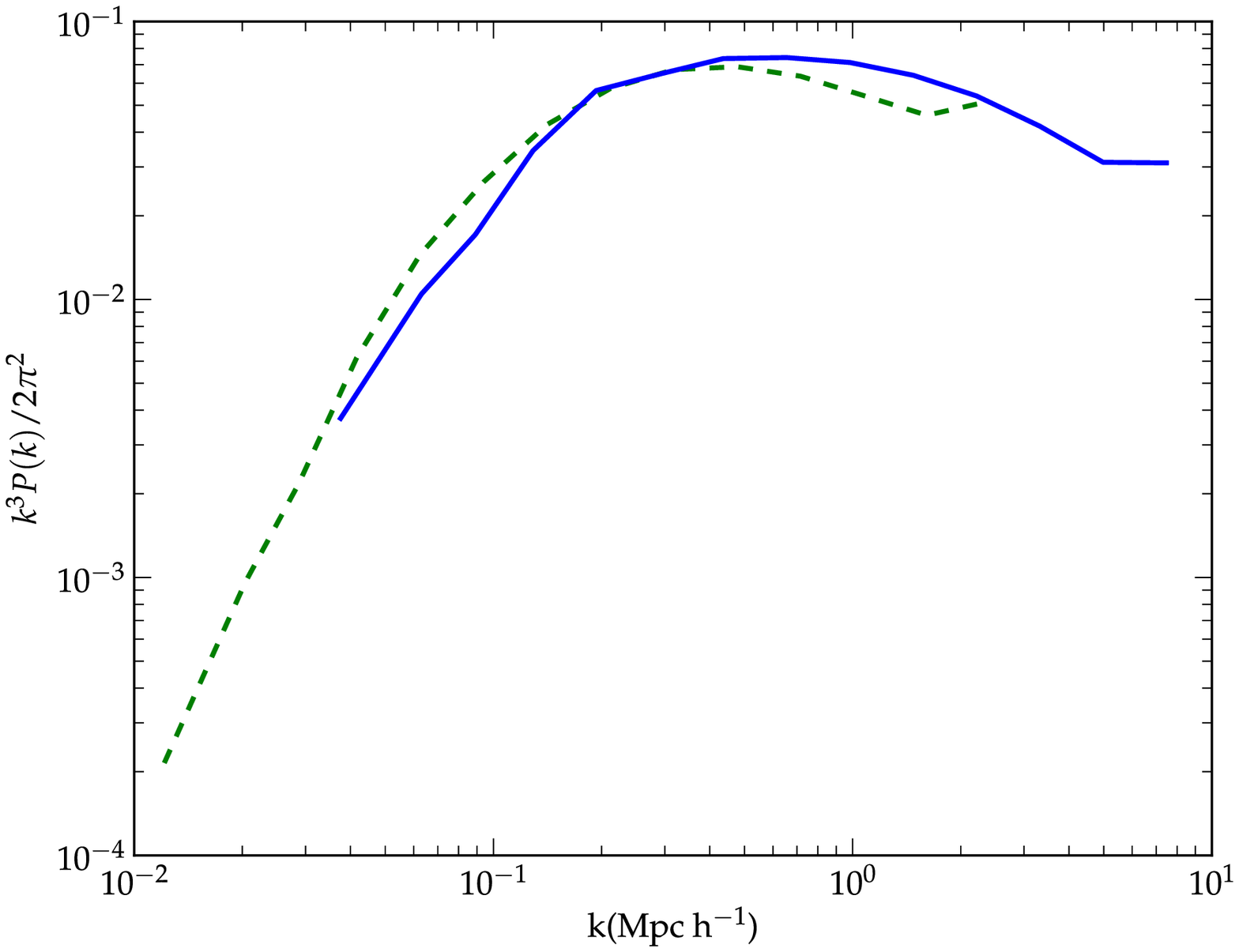}}
\caption{Power spectrum of the ionization field for the very large volume simulation (green dashed curves), compared to the same result for the L=300 Mpc simulation. Both plots are for z=9 but using different efficiency parameters ($\zeta$=5.0 and $\zeta$=14.0) so that $\bar{x}_i=0.1$ (left) and $\bar{x}_i=0.55$ (right), respectively.}
%\vspace{0.5cm}
\label{ps_xi_large}
\end{figure*}

Encouraged by this result, we proceed to derive the ionization bubbles using the same excursion-set formalism as in Section 2. For this, we have also applied the Zel'dovich approximation for the density and halo field boxes. In order to keep a moderate computational time, we used again a N=600$^3$ box for determining the ionization field, which resulted in a cell size of about 1.5 Mpc. This required the need to identify smaller ionized regions ($\sim$0.5 Mpc) which was done using the same technique introduced in Section 2, with the halo catalog being used to identify the ionized volume within each non-ionized cell.
The results for the power spectrum of the ionization field are shown in figure \ref{ps_xi_large} compared to those obtained with the full excursion-set method with L=300 Mpc and for two different efficiency parameters at z=9. The curves seem to agree reasonable well showing convergence with the 300 Mpc box, although if we go to even larger ionization fractions we still find slightly larger bubbles in the 1Gpc box (the largest bubble for the 1Gpc box with $x_i=0.55$ has a size of 20 Mpc). Note that the bubble size will also be affected by absorption systems (e.g. \citealt{alvarez10}).
%%%%%%%%%%%%%%%%%%%

\begin{figure*}
\centerline{\hspace{0.2cm}\includegraphics[scale=0.5]{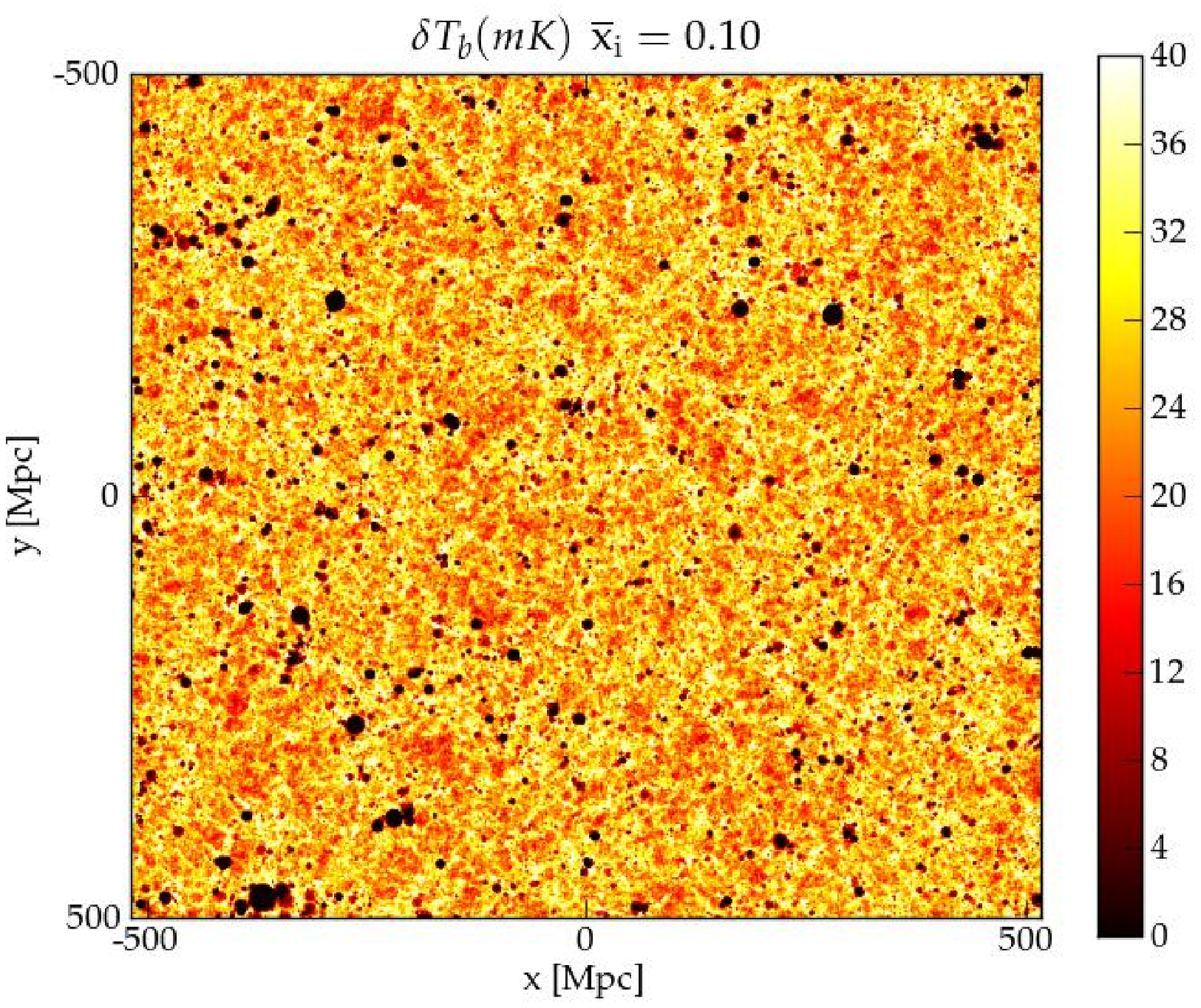}\hspace{-1.2cm}  \includegraphics[scale=0.5]{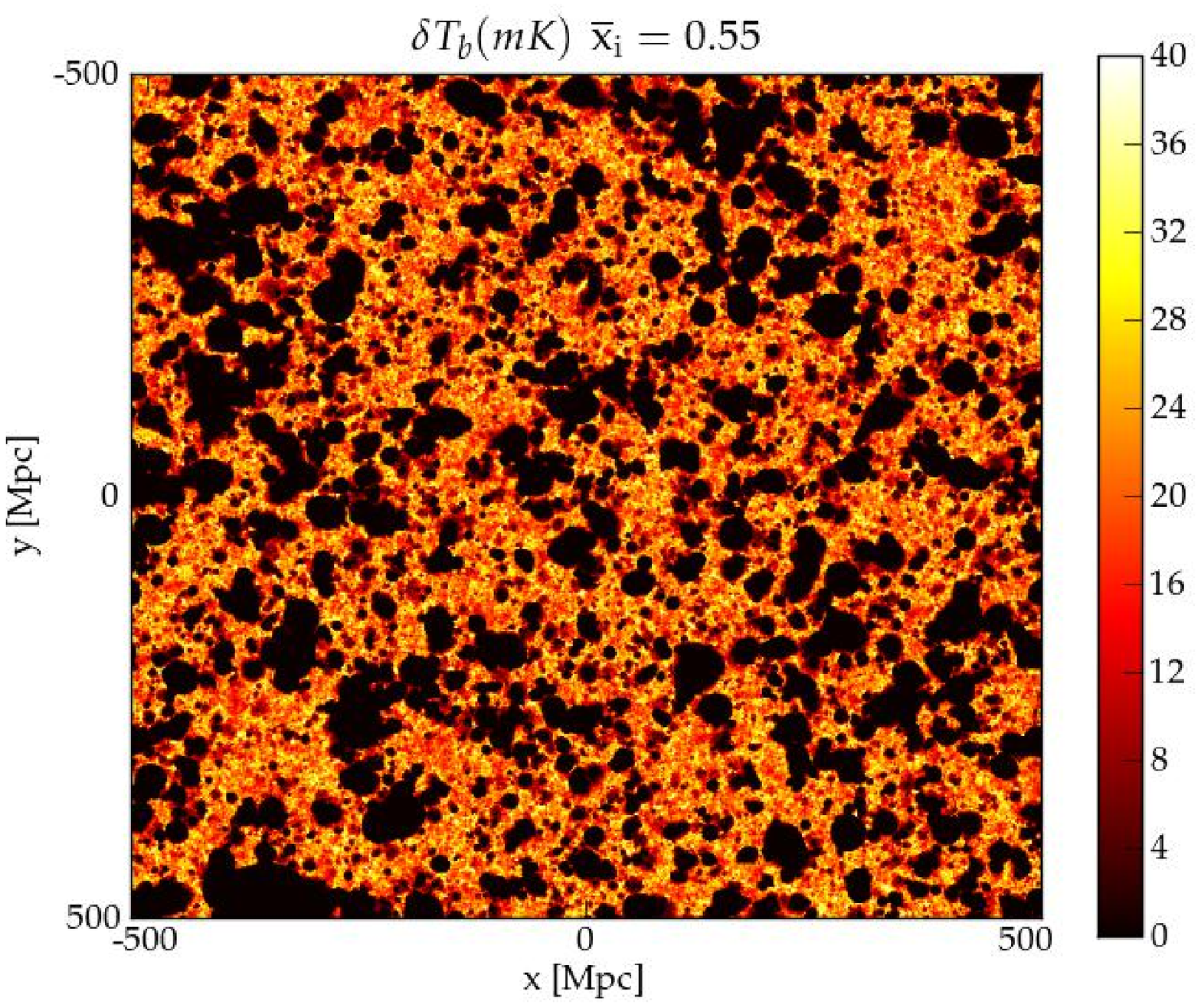}}
\caption{The temperature maps for the L=1000 Mpc simulation for two different ionization states at z=9 (as in figure \ref{ps_xi_large}).}
%\vspace{0.5cm}
\label{large_maps}
\end{figure*}

To illustrate the application of this method we show the final brightness temperature maps for the expected 21-cm signal in the L=1000 Mpc simulation (fig. \ref{large_maps}). Note that we are considering partially ionized cells in this simulation, so even the large patches of emission in the figure have a reasonable amount of ionizing sources, although this is difficult to see given the small size of each ``pixel''.
%%%%%%%%%%%%%%%%%

\section{Summary}

We presented a semi-numerical method capable of quickly generating end-to-end simulations of the 21 cm signal even at the high redshifts where the spin temperature is non-negligible. The algorithm allows to generate brightness temperature boxes with very large volumes, e.g. (1000 Mpc)$^3$, crucial to properly simulate the field of view of the next generation of radio-telescopes, without sacrificing the speed or requiring unfeasible computer resources\footnote{During the review process of this paper another algorithm was presented by \citet{mesinger10}.}.

The corresponding code (SimFast21) implements the following prescription:
\begin{itemize}
\item
Monte Carlo generation of the dark matter density and velocity fields at z=0 assuming Gaussian distribution functions.
\item
Determination of the halo catalog using an excursion-set formalism on the linear density field at any given redshift. 
For large volumes a biased Poisson sampling is used to introduce small mass halos.  
\item
Adjustment of the halo and dark matter locations using the Zel'dovich approximation.  
\item
Simulation of $x_i({\bf x},z)$ from FFTs of the previous boxes.
\item
Extraction of the star formation rate density from the halo mass distribution.
\item
Extraction of gas temperature fluctuations $T_K({\bf x},z)$ .
\item
Calculation of the Ly$_\alpha$ coupling and collisional coupling parameters.
\item
Determination of the 21 cm brightness temperature boxes using the above quantities.
\end {itemize}
The computational time required for the above steps depends mainly on the volume to be covered (our bubble filtering procedure is now pratically independent of the ionization fraction due to the FFT based algorithm we are now using). This basically affects the halo determination running time, since larger volumes imply a larger number of halos and operations for overlap checking. 
Nevertheless, the typical time for the computation of the above steps is about 2 hours for the simulation with L=1000 Mpc, which can be considered remarkably fast. We are expecting that further optimization of the code can still reduce this time.

Although much faster than hydrodynamical numerical simulations, our analysis shows that relevant quantities such as the halo mass function, the halo mass power spectrum, the star formation rate or the ionization fraction power spectrum are all consistent with the numerical simulations with which we compared our results to. In particular, the ionization fraction power spectrum is similar to the one obtained with radiative transfer codes (RT) when comparing to the same volume. We still find some differences on the largest scales when comparing to simulations with relatively small volumes (143 Mpc), which we can relate to the production of larger HII bubbles. These differences become much smaller as we increase the simulation volume and the change of size in the largest bubbles becomes less important when compared to the overall size of the simulation.

The dependence on the astrophysical parameters of the simulation was encoded in three functions: the ionization efficiency, $\zeta$, the Ly$_\alpha$ spectral distribution function of the sources, $\epsilon_\alpha$ and the X-ray spectral distribution function, $\epsilon_X$. These functions can be easily changed for a model of our choice and the code can then quickly generate new simulations of the signal (even faster if we keep the same cosmology). By combining all the unknowns into a physically meaningful small set of parameters we can easily probe the huge intrinsic parameter space available at high redshifts probed by 21 cm observations.
This versatility is included in our algorithm. The possibility
to generate reionization simulations from scratch without the need for supercomputers is a great advantage allowing to experiment with different models. 
Moreover, this allows room for
improvement through calibration with full numerical simulations. The code can be a useful tool to generate sky models for future 21 cm experiments, important to test the observation pipeline, study how the foregrounds affect the observations and can be separated given a noise model,
and to develop optimal estimators for signal extraction. The code will be released for public use in due course and we welcome the community participation in its updating and upgrading as well as development of additional applications on observational probes of reionization beyond the 21 cm observations.

\section*{Acknowledgments}

We would like to thank R. Cen and H. Trac for providing the data from their simulation.
This work was partially supported by FCT-Portugal under grants PTDC/FIS/66825/2006 and PTDC/FIS/100170/2008.
The work was also supported by the European Community Framework Programme 6, Square Kilometre Array Design Studies (SKADS), contract no 011938.

\bibliography{simfast21}
\bibliographystyle{mn2e}

\label{lastpage}

\end{document}